\documentclass[12pt]{iopart}
\usepackage{graphicx}

\usepackage{cite}
\usepackage{wrapfig}

\begin{document}


\title{Scroll ring chimera states in oscillatory networks}

\author{Volodymyr Maistrenko$^{1}$,  Oleksandr Sudakov$^{1,2}$, Ievgen Sliusar$^{1,2}$}

\address{$^1$Scientific Centre for Medical and Biotechnical Research, National Academy of Sciences of Ukraine, Volodymyrska Str. 54, 01030 Kyiv, Ukraine 
\\

$^2$ Taras Shevchenko National University of Kyiv, Volodymyrska Str. 60, 01030 Kyiv, Ukraine}
\ead{maistren@nas.gov.ua}

\begin{abstract}
We report the appearance of a scroll ring and scroll toroid chimera states from  the proposed initial conditions for the Kuramoto model of coupled phase oscillators in the 3D grid topology with inertia. The proposed  initial conditions provide an opportunity to obtain as single as well as multiple scroll ring and toroid chimeras with different major and minor diameters. 
We analyze their properties and  demonstrate, in particular, the patterns of coherent, partially coherent, and incoherent  scroll ring chimera states with different  structures of filaments and chaotic oscillators.  Those patterns
can coexist  with solitary states and solitary patterns in the oscillatory networks. 
\end{abstract}


\maketitle

 Scroll rings appear in the models  of  various real physical systems and are among the most paradigmatic
examples of spatio-temporal self-organizing structures in
excitable media.  Scroll rings occur in three-dimensional media and are formed from a scroll wave. In general, scroll rings are three-dimensional spiral
waves rotating around closed one-dimensional space curves.
A scroll wave is usually characterized by its filament that can be considered as the
line connecting the rotation centers of the spirals in the two-dimensional
cross-sections of the scroll wave. Then this filament can 
be closed into a ring  structure called the scroll ring  \cite{w1973}.

Scroll rings have been observed in a variety of systems including the chemical
Belousov--Zhabotinsky reaction \cite{w1972}, in a fibrillating cardiac
tissue \cite{mpp1984}, etc. 
The existence of a scroll ring in the media of various nature was reported as the results of numerous simulations and experiments in the fields of physics, chemistry, biology, etc.
(see, e.g.,   \cite{ws1983,pr1987,w2001,cf2008,qhgw2014,bb2014,tes2015}).

Chimera states  \cite{k2002,kb2002,as2004} as a phenomenon of coexistence of coherent and incoherent patterns in the arrays of nonlocally coupled
oscillators were investigated in a wide range of systems. 
 For these special spatio-temporal patterns,  some network's
elements oscillate synchronously with unique frequency, and the others behave themselves asynchronously.
A number of articles are devoted to the theoretical and experimental studies of chimera states. Most of them deal with the models of one- and  two-dimensional  networks of oscillators \cite{pjaswbp2021}.

Recently, the chimera states in a three-dimensional grid topology were investigated  within various models of coupled phase oscillators  \cite{msom2015,ld2016,msom2017,khp2018,ok2019,kbgl2019,mso2020}.
 The first report  of  a stable scroll ring chimera state  was done in 2020 in  \cite{mso2020} for the Kuramoto model of coupled phase oscillators in a 3D grid topology with inertia. This pattern was  obtained from random initial conditions for the following equation:

\begin{eqnarray}
m\ddot{\varphi}_{ijk} + \epsilon \dot{\varphi}_{ijk}  = 
 \frac{\mu}{|B_{P}(i,j,k)|} \sum\limits_{(i^{\prime},j^{\prime},k^{\prime})\in B_{P}(i,j,k) }\sin(\varphi_{i^{\prime}j^{\prime}k^{\prime}} - \varphi_{ijk}- \alpha),
\end{eqnarray}
where   $i,j,k = 1, ... , N$, $\varphi_{ijk}$ are phase variables, and the indices  $i,j,k$ are periodic modulo $N$.   The coupling  is  assumed long-ranged and isotropic:  each oscillator $\varphi_{ijk}$ is coupled with equal strength $\mu$ to all its nearest neighbors $\varphi_{i^{\prime}j^{\prime}k^{\prime}}$  in  a ball of radius $P$, i.e., to those falling in the neighborhood 
\vspace*{-0.1cm}
$$  B_{P}(i,j,k):=\{ (i^{\prime},j^{\prime},k^{\prime}){:} (i^{\prime}-i)^{2}+(j^{\prime}-j)^{2}+(k^{\prime}-k)^{2}\le P^{2}\},$$
where the distances  $|i^{\prime}-i|$, $|j^{\prime}-j|$  and  $|k^{\prime}-k|$ 
are calculated regarding the periodic boundary conditions of the network.
$|B_{P}(i,j,k)|$ denotes the cardinality of $B_{P}(i,j,k)$.
The phase lag parameter $\alpha$ is selected from  the segment $[0, \pi/2]$.  
The  relative coupling radius $r=P/N$ varies from $1/N$ (local coupling) to $0.5$ (close to the global coupling).

The parameter $\mu$ is the oscillator coupling strength, and $\epsilon$ is the damping  coefficient.
 The parameter $m$ is the mass. In the case $m=0,$ Eq. (1) is transformed into the 3D Kuramoto model  without inertia.
We put $m=1$ without any loss of generality.

In the case of scroll ring chimera states, a filament that consists of oscillators and connects scroll waves'  
rotation centers has a shape of  ring or  toroid  with different major and minor diameters  in the 3D oscillatory network.  

In the present paper, we propose the  initial conditions  for the generation   of scroll ring chimera states in the Kuramoto model of coupled phase oscillators in the 3D grid topology with inertia (1). They can 
generate a variety of scroll ring  and scroll toroid chimeras with different shapes and inner structures of their filaments.
We study  the properties of scroll ring chimeras and will demonstrate  that they  can be coherent, partially coherent, or incoherent with a hyper-chaotic behavior  
characterized by a number of positive Lyapunov
exponents.

Scroll ring chimeras in the Kuramoto model (1) can be surrounded by solitary oscillators and can coexist with other patterns in the 3D oscillatory network. The
solitary state behavior means that some number of oscillators start to rotate
with a different time-averaged frequency as compared to the synchronized oscillators\ \cite{JMK2015,JBLDKM2018,msm2020,mso2020}.

Numerical simulations were performed on the base of the Runge--Kutta solver DOPRI5 on  the Ukrainian Grid Infrastructure  with graphics processing units  \cite{sls2011,scm2017}. 
 In total, more than 10000 network trajectories were computed and analyzed for system (1) with $N = 100$ and $200$ (1 and 8 millions of oscillators, respectively).

 To obtain the scroll ring chimeras in system (1), we propose two types of the following  initial conditions:

\vspace*{-0.6cm}
\begin{eqnarray}
\varphi_{xyz} = -\theta \exp{ \left(-\frac{2r_{rot}}{D} \right)}, \\
\varphi_{xyz} = -\theta \exp{\left(-\frac{\vert r_{xy} \vert +\vert 2z-1\vert}{2D}\right)}, 
\end{eqnarray}
\vspace*{-0.4cm}
\begin{displaymath}
\dot{\varphi}_{xyz} =  \left\{ \begin{array}{ll}
  \omega_{max}  \ \textrm{if} \ (2x-d_{x})^{2}+(2y-d_{y})^{2}+(2z-1)^{2} \leq d^{2}, \\
\omega_{min} \  \textrm{if} \  (2x-d_{x})^{2}+(2y-d_{y})^{2}+(2z-1)^{2} > d^{2},
  \end{array} \right.
\end{displaymath}
\vspace*{-0.2cm}
where 
\begin{eqnarray}
r_{rot} = \sqrt{(r_{xy})^2 + (z-0.5)^{2}}, \nonumber  \
r_{xy} = (x-0.5) \cos(\psi) + (y-0.5) \sin(\psi) + \frac{D}{2}, \nonumber  \\
d_{x} = 1 + D \cos(\psi), 
d_{y} = 1+ D \sin(\psi),  \nonumber \
\theta = {\textrm{atan2}} ({z-0.5},r_{xy}), \nonumber  \\
\psi =  {\textrm{atan2}} ({y-0.5, x-0.5}),  \nonumber  \
x=i/N,  y=j/N,  z=k/N.  \nonumber 
\end{eqnarray}

The parameters $D, d \in (0, 1]$.  $x, y, z \in  [1/N, 1], \psi \in [- \pi, \pi]$.

\begin{figure}[ht!]
\centering  
\includegraphics[width=0.4\linewidth]{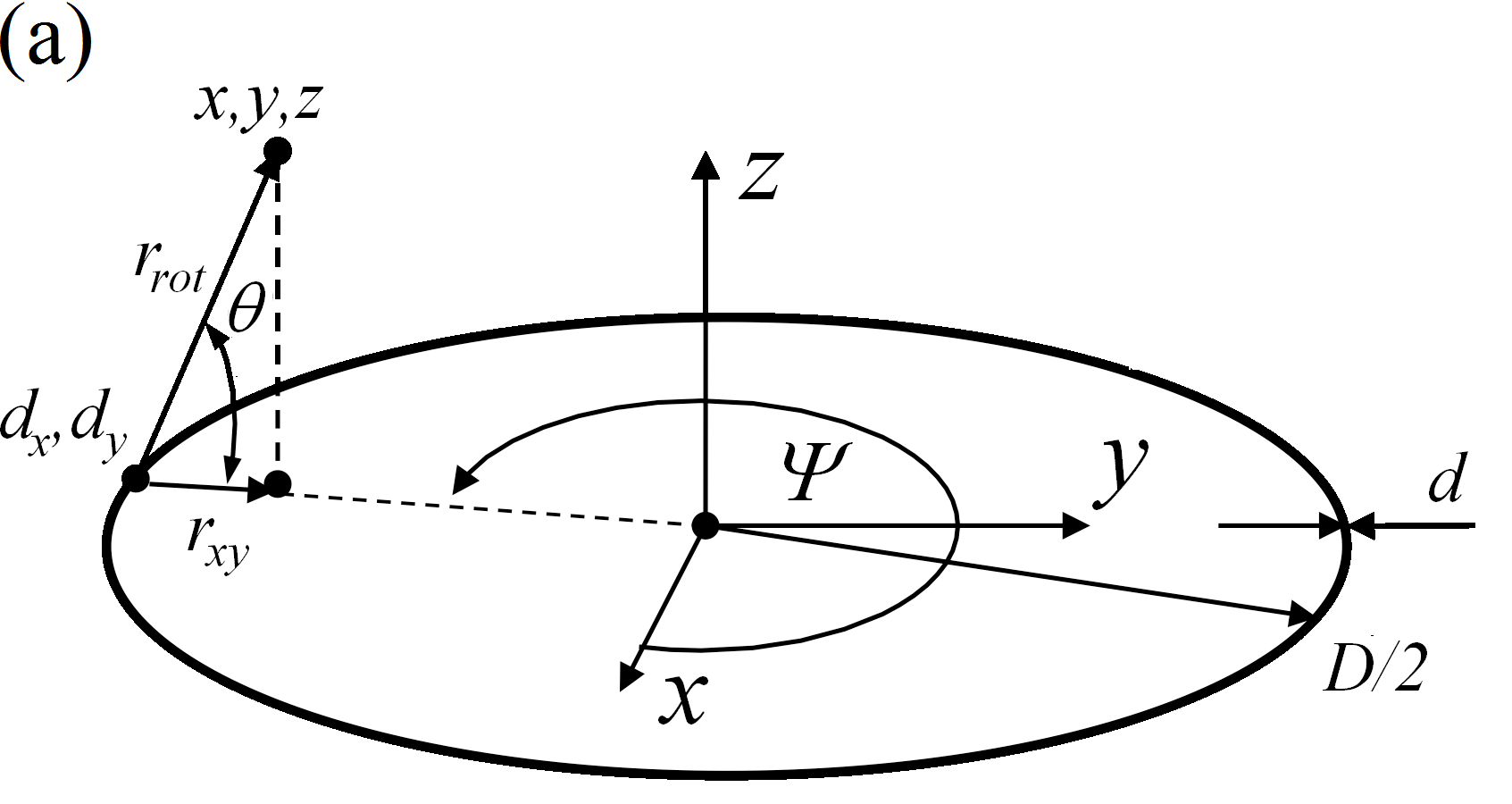}  \hspace*{0.2cm} 
 \includegraphics[width=0.2\linewidth]{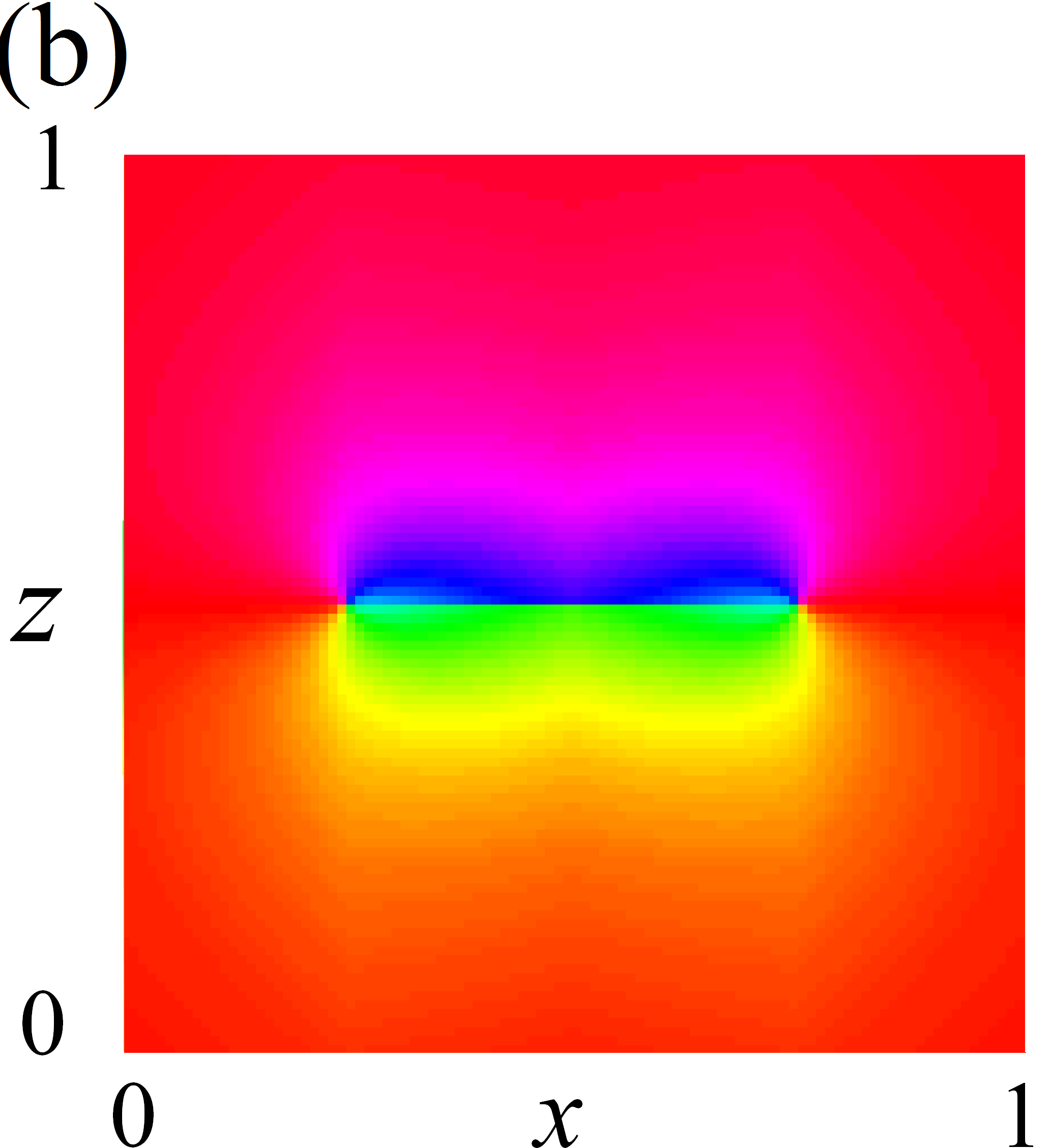}  \  \   \hspace*{0.2cm} 
  \includegraphics[width=0.2\linewidth]{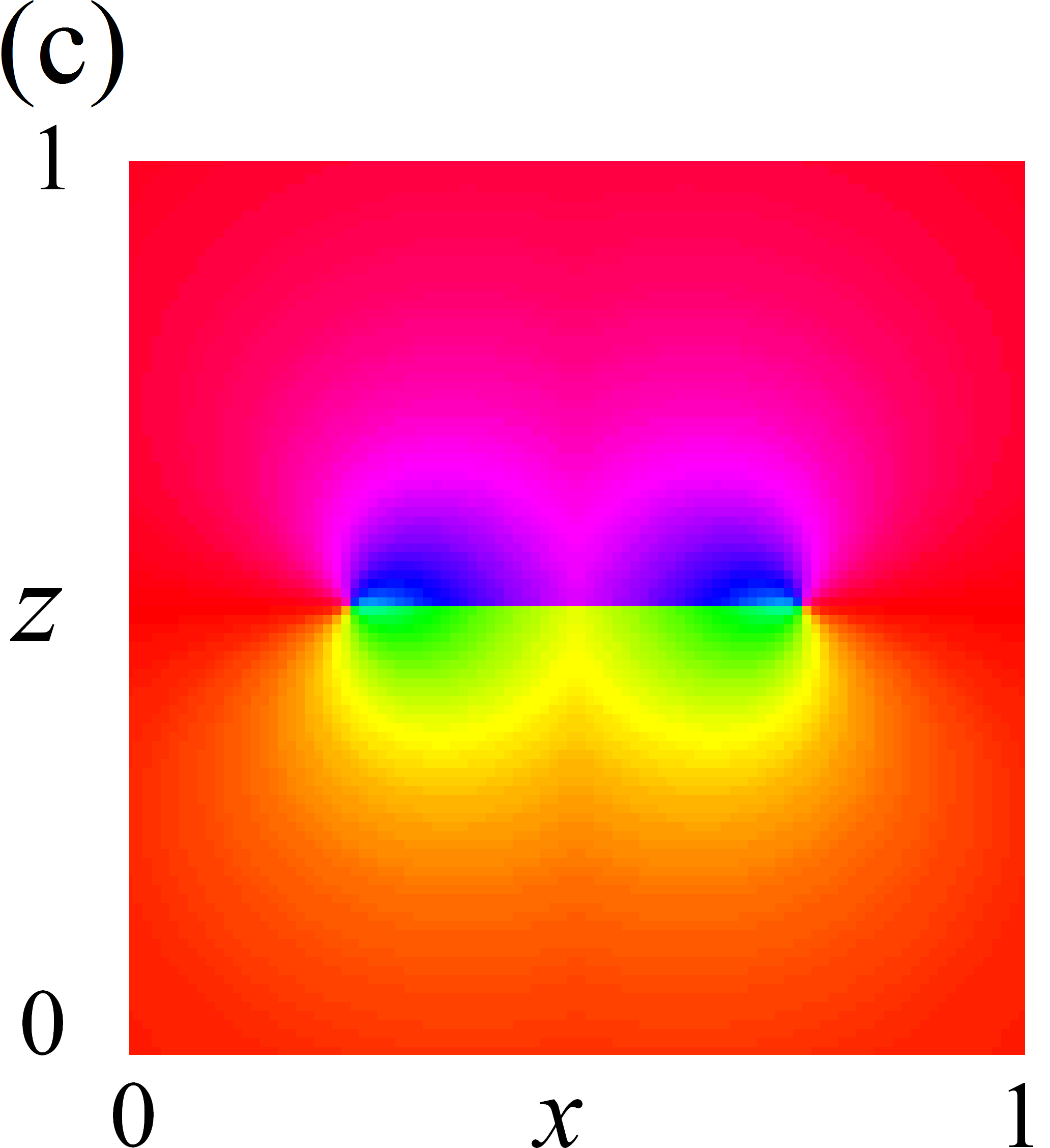}  \  \   
  \includegraphics[width=0.035\linewidth]{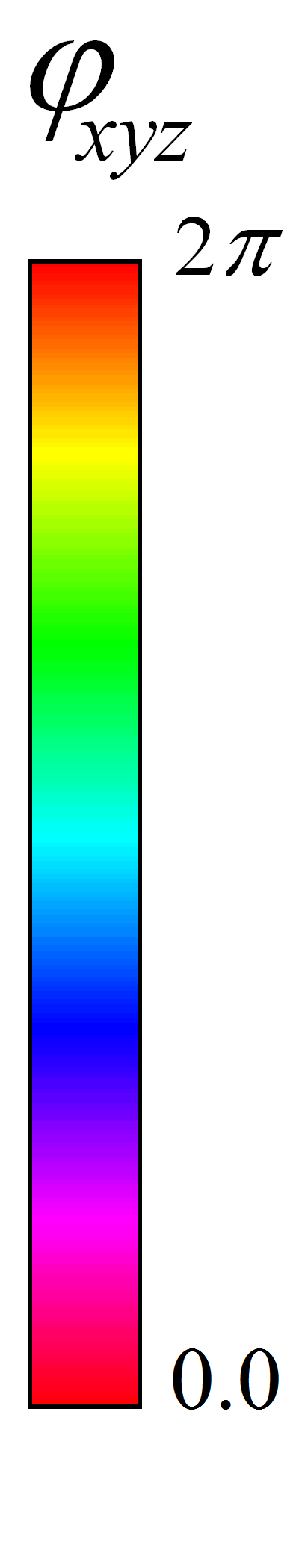}
\hspace*{-2.8cm}
\begin{minipage}{1.22\textwidth}
\caption {Schematic view of the construction of initial conditions (a).  Phase cross-sections  of initial conditions $\varphi_{xyz}$ modulo $2 \pi$ along y=0.5: (b) - type  (2), (c) - type (3). 
Parameters  $D=0.5, d=0.04, N=100$. }
\end{minipage}
\end{figure}

 The  initial conditions  (2), (3) describe the rotation of the phase $\varphi_{xyz}$ around all axes tangent to the circle with diameter
$D$ in the plane $z = 0.5$. The schematic illustration of their construction is presented in Fig. 1(a). To satisfy the boundary conditions, the phase exponentially damps from the rotation center to the boundary. The fast damping to zero provides the same phase at large distances from the
rotation center. The damping laws are different for (2) and (3). The analytic expression (2) yields a slower
damping in the $(x, y)$ plane and a faster damping in the  $z$ direction. The analytic expression (3) describes the symmetric
damping around the rotation axis.
A single scroll ring without phase damping satisfies periodic boundary conditions in the plane
$z = 0.5$.  The symmetric phase damping (3) is considered as the most
simple case of damping law for the scroll rings. Figures 1(b) and 1(c) demonstrate the difference between the initial conditions (2) and (3) by the presentation of the cross-sections of the phase $\varphi_{xyz}$ modulo $2 \pi$ along $y=0.5$ at the parameter values  $D=0.5, d=0.04, N=100$.

The value of the parameter $\omega_{max}$ is the  maximal frequency of chimera's oscillators,  $\omega_{min}$ is the frequency of the rest synchronized oscillators for any chimera states  (not necessarily a scroll ring) obtained for fixed parameter values   $\alpha, \epsilon, \mu, r, N$ of model (1), which  can be obtained from the random initial condition at these parameter values. These parameters must be determined before starting the simulation procedure.

\begin{wrapfigure}{r}{0.48\textwidth}
\centering  
\includegraphics[width=1.0\linewidth]{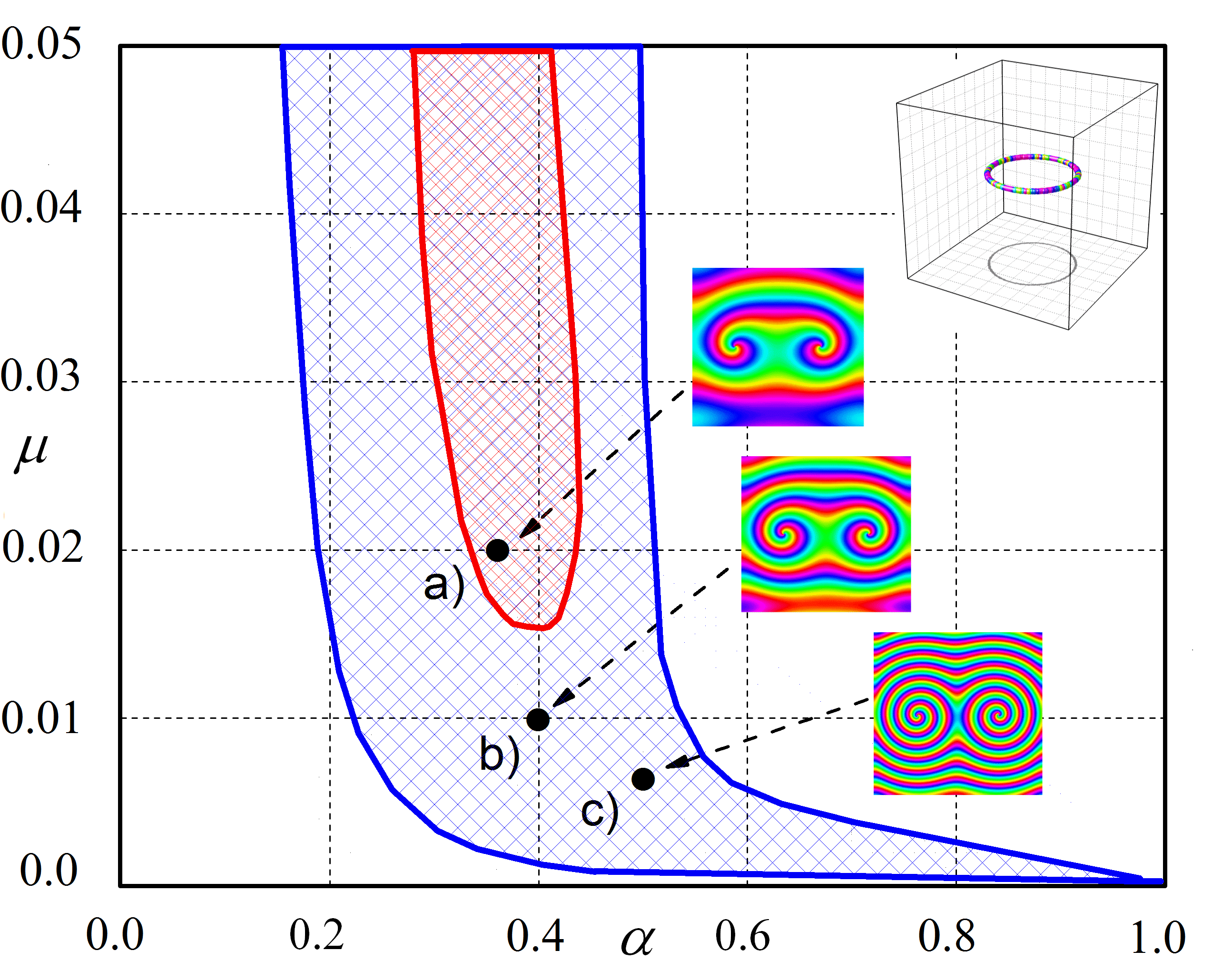} 
\hspace*{-2.4cm}
\begin{minipage}{0.62\textwidth}
\caption{Stability regions  for scroll ring chimeras  in the parameter planes $(\alpha, \mu)$. 
Blue - scroll rings with the incoherent or partially coherent inner part, red - coherent rings. Snapshots of typical scroll ring chimeras and cross-sections of their phases $\varphi_{xyz}$ along y=0.5 are shown in the insets. Parameters $\epsilon=0.05, r=0.01, N=200$.  }
\end{minipage}
\end{wrapfigure}

In this way, the initial conditions (2), (3) guarantee the generation of single scroll ring or toroid chimeras, as well as  multiple scroll rings or other scroll wave chimeras.  The scroll toroid chimeras  must have minor diameter more than $1/N$.  

Nevertheless, although the minor diameter of resulting scroll ring chimeras is determined largely by the parameter $d$,  the  value of the relative coupling radius $r=P/N$ has also a great influence on it. 

Before starting the simulation of scroll ring and scroll toroid chimeras with expected major and minor diameters, we must preliminary calculate the parameter domains of model (1), where the patterns exist and are stable, by  using proposed initial conditions.
 Such regions for the scroll ring chimeras in the  parameter plane $(\alpha, \mu)$ at $\epsilon=0.05,  r=0.01,  N=200$ is presented in Fig. 2.
 These regions are similar to the stability regions for scroll ring chimeras obtained from the random initial conditions, but for another value of $r=0.04$ (see Fig. 3(a)) in   \cite{mso2020}).

Our simulation shows that the scroll ring chimera states exist for any infinitely small coupling strength $\mu>0$. Crossing the left and left bottom sides of the scroll ring stability region, all oscillators are synchronized. After crossing the right side of the region, the rings are destroyed with the generation of the chaotic oscillatory mode or their total synchronization.

The examples of typical scroll ring chimeras and their cross-sections  in the insets of Fig. 2 at parameter point a) - c) demonstrate that the scroll waves are more densely twisted around the scroll ring chimeras with decreasing the parameter $\mu$.

\begin{figure}[ht!]
 \center{\includegraphics[width=0.44\linewidth]{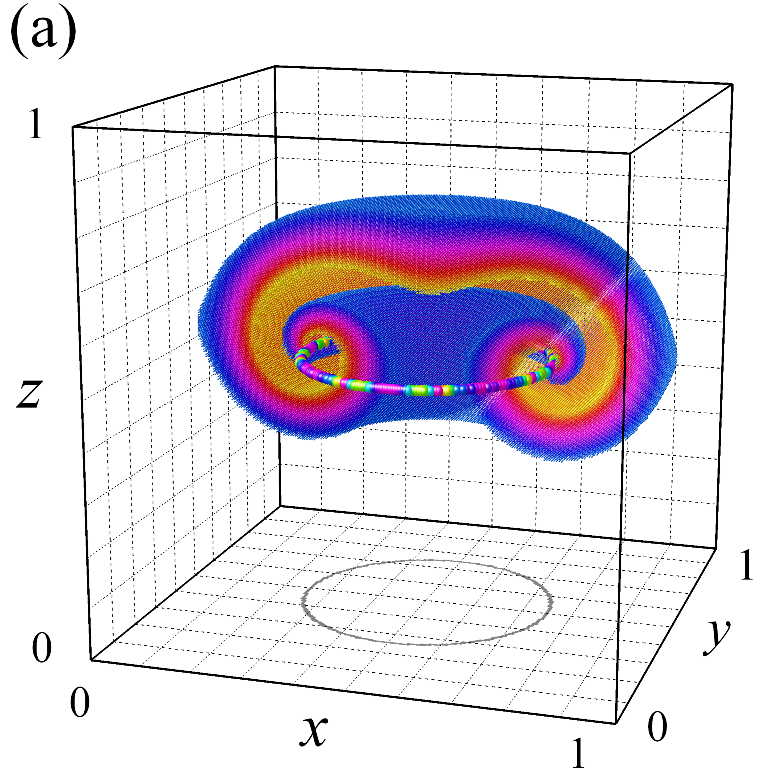}  \  \   
  \includegraphics[width=0.44\linewidth]{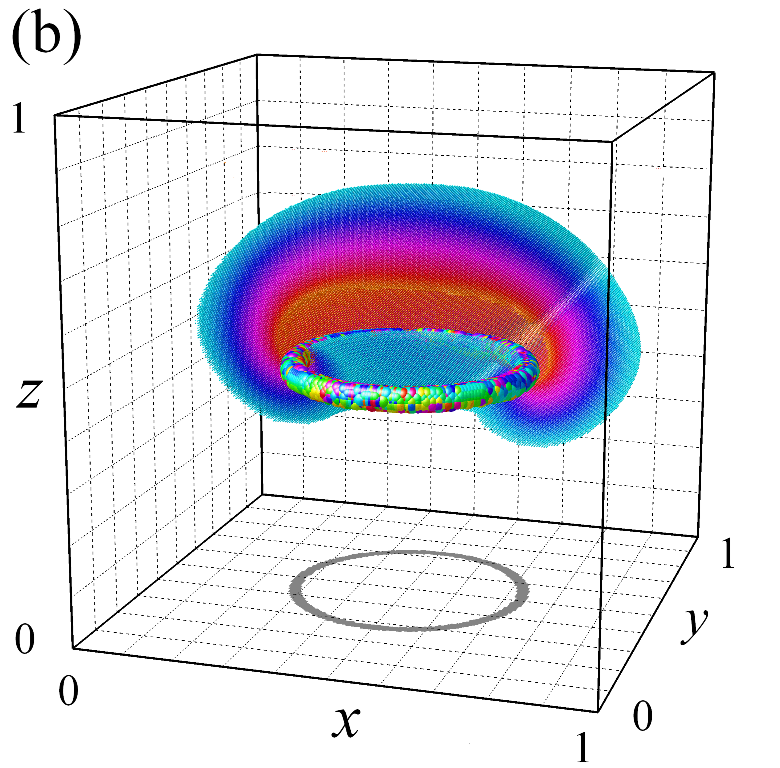}  \  \   
  \includegraphics[width=0.05\linewidth]{Faza-insert-F4.png}} 
\hspace*{-2.8cm}
\begin{minipage}{1.15\textwidth}
 \caption{  Phase snapshots of  scroll ring and scroll toroid chimeras obtained from the initial conditions (2): (a) - scroll ring chimera ($\mu=0.016, r=0.01, d=0.005$),  (b) -  scroll  toroid chimera ($\mu=0.01, r=0.03, d=0.08$). Common parameters $\alpha=0.4, \epsilon=0.05, D=0.5, N=200$. Simulation
time $t = 10^{5}$. Half of the wave has been removed to permit
the view of the scroll chimeras.
 Coordinates  $x=i/N,  y=j/N,  z=k/N$. The space–time dynamics of the scroll rind and scroll toroid chimera states are illustrated by videos in supplemental data. } 
\end{minipage}
\label{fig:3}
\end{figure}

Figure 3 illustrates the results of  simulation with the use of the initial condition (2) for scroll ring chimera (a) and  for scroll toroid chimera (b) with major diameter of $0.5$. In the case of 
scroll toroid chimera, its minor diameter is far more than $1/N$, which clearly seen in Fig. 3(b). In particular, the projection of the pattern onto the $(x, y)$ plane confirms it. 
Here, we take  $\omega_{max}=0.06$, $\omega_{min}=-0.19$.
In these figures, a half of the wave has been removed to permit one
to better see the chimeras.

\begin{figure}[th!]
 \center{
\includegraphics[width=0.29\linewidth]{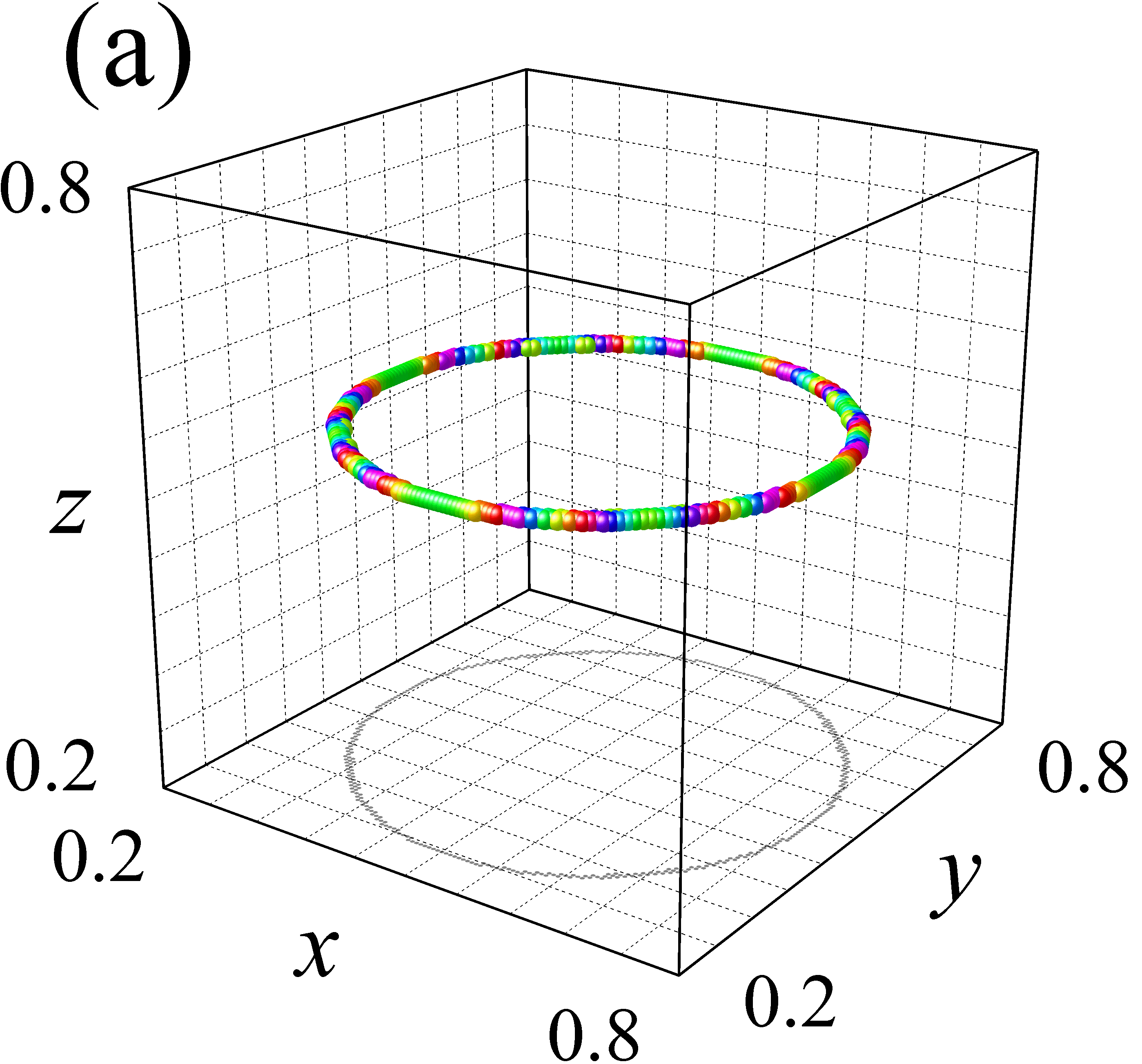}  \ \ 
\includegraphics[width=0.29\linewidth]{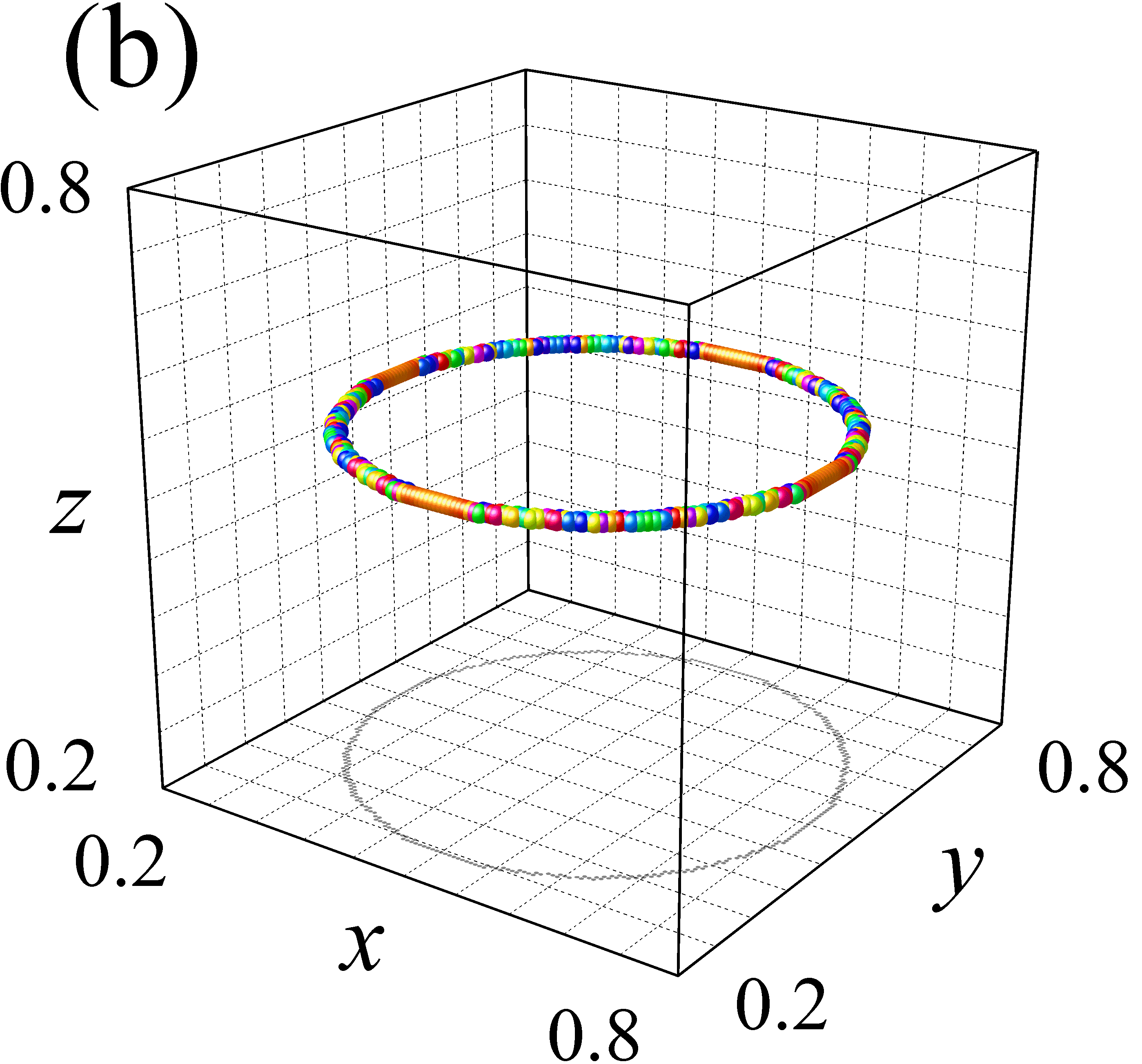}  \ \ 
\includegraphics[width=0.29\linewidth]{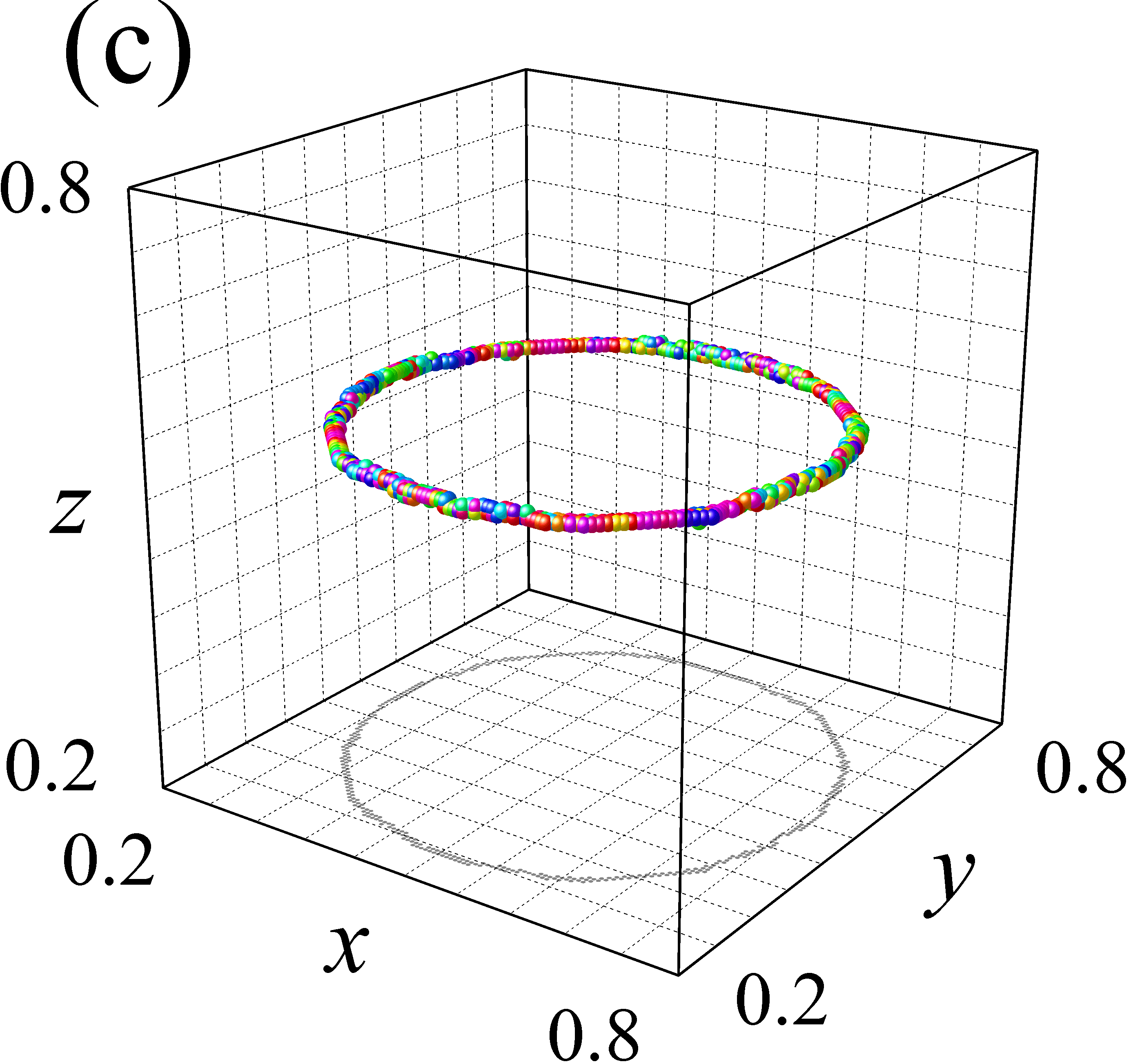}  \ \   
   \includegraphics[width=0.04\linewidth]{Faza-insert-F4.png}  

\vspace*{0.2cm}
\includegraphics[width=0.29\linewidth]{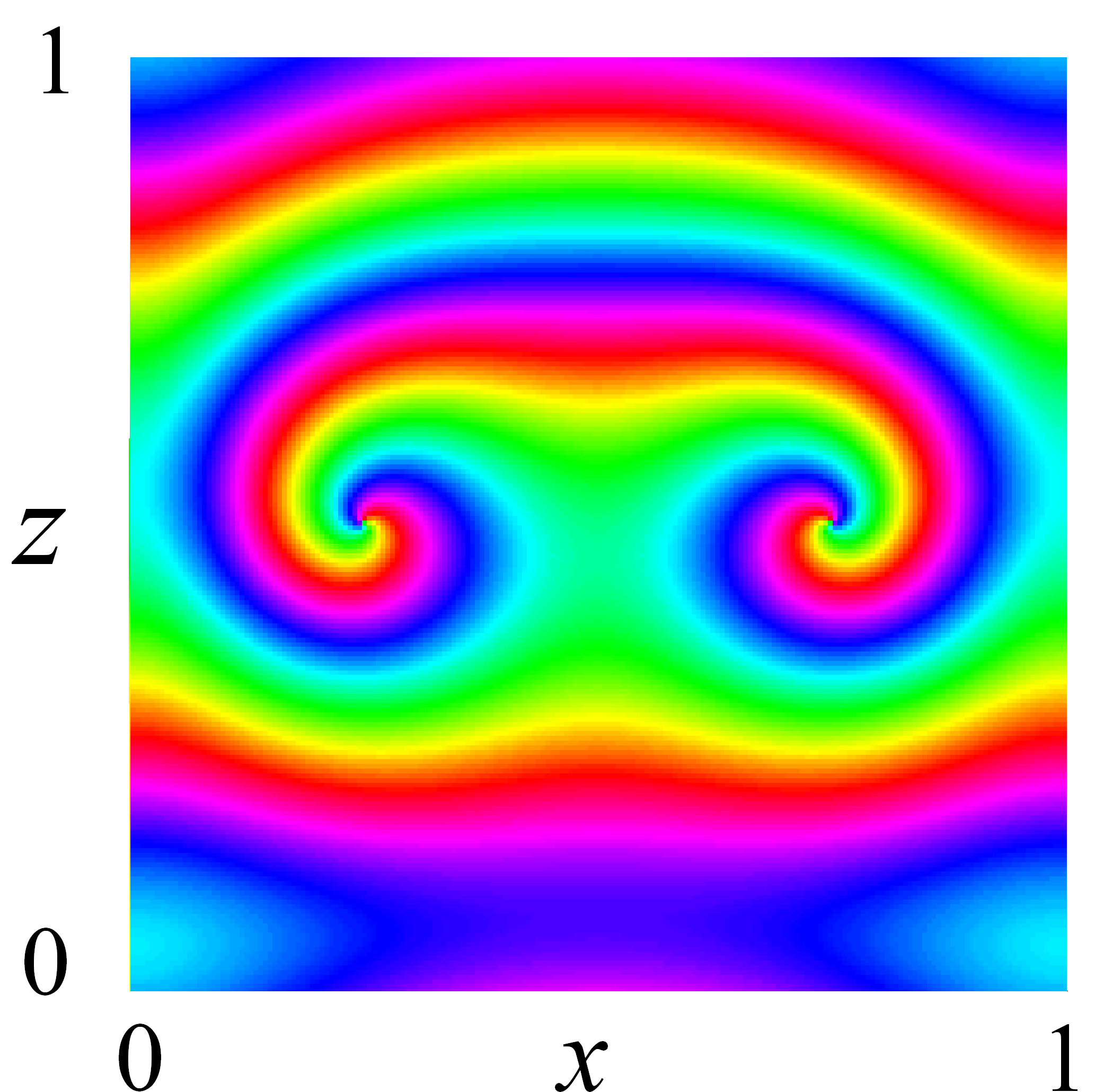}  \ \ 
\includegraphics[width=0.29\linewidth]{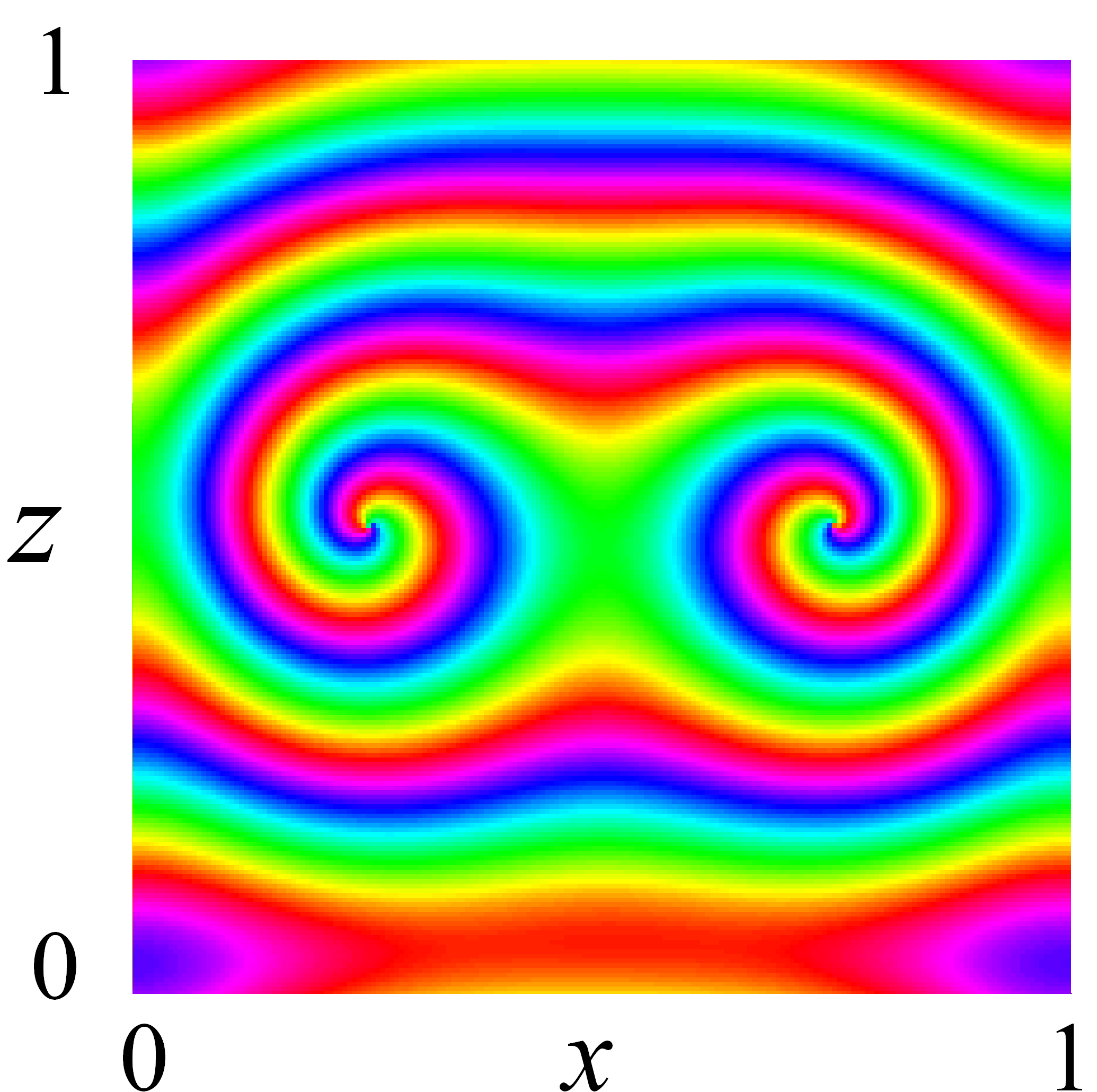}  \ \ 
\includegraphics[width=0.29\linewidth]{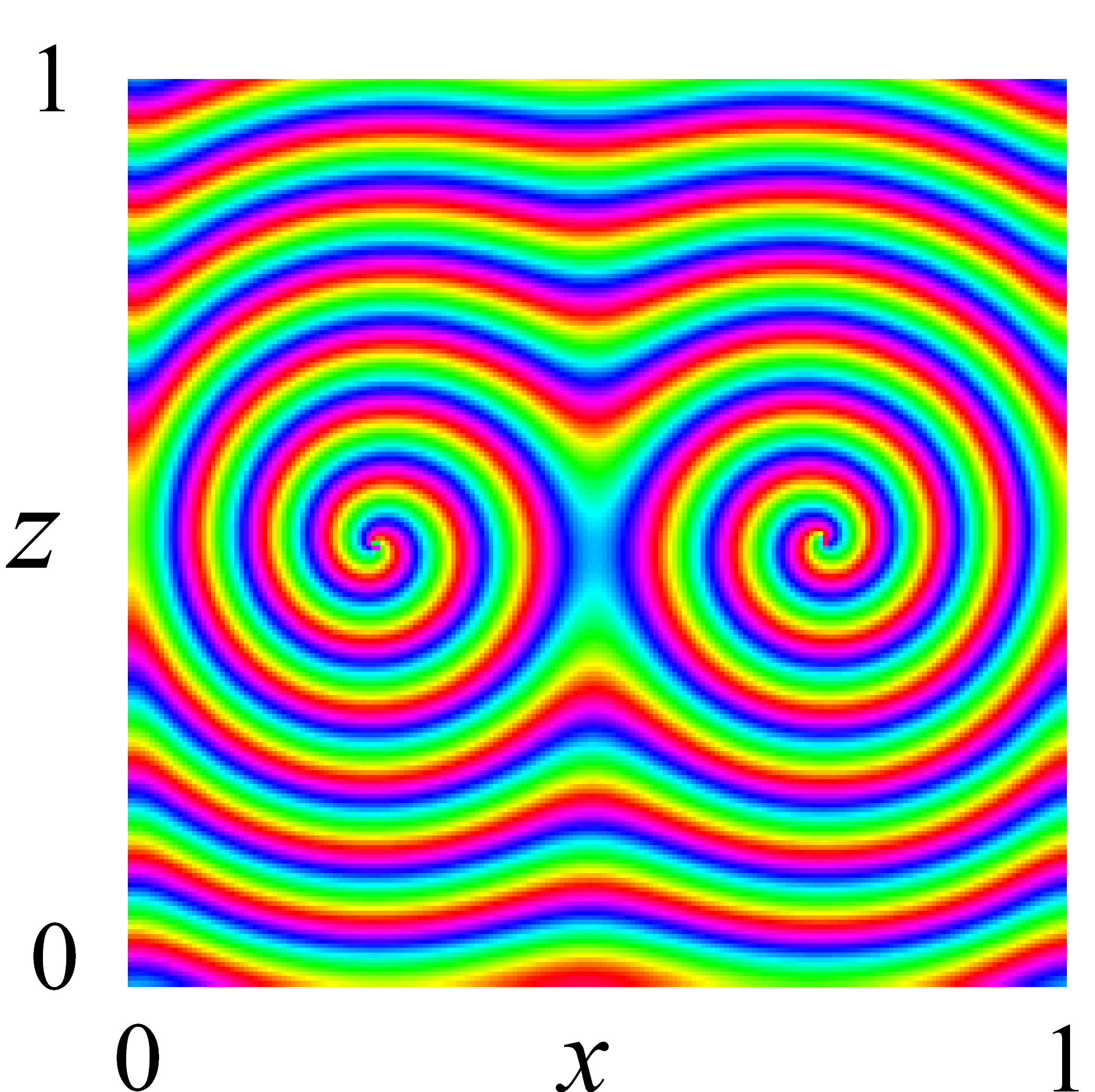}  \ \ 
   \includegraphics[width=0.04\linewidth]{Faza-insert-F4.png}  \ 

\vspace*{0.4cm}
  \includegraphics[width=0.29\linewidth]{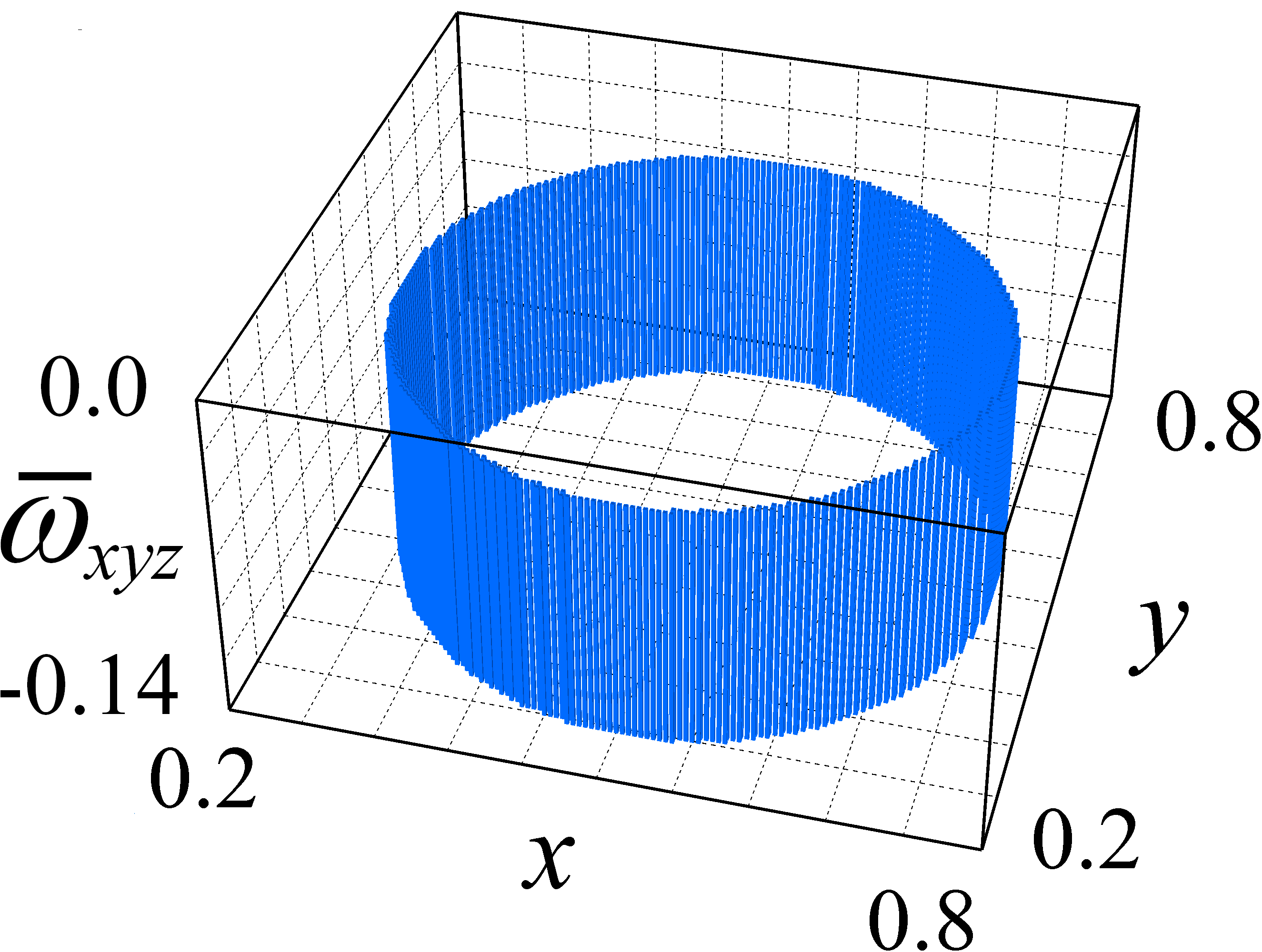} \ \ 
  \includegraphics[width=0.29\linewidth]{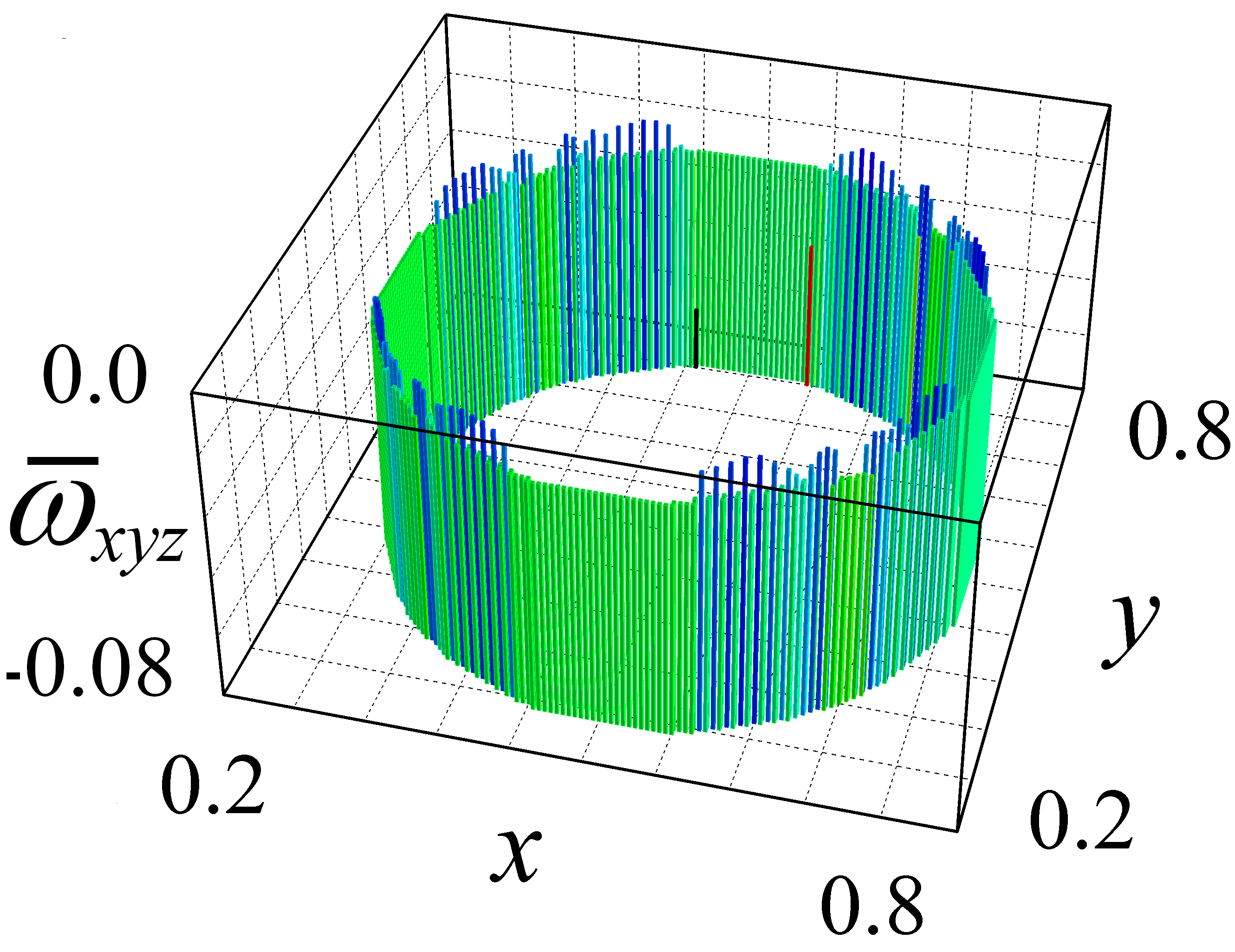}  \ \
  \includegraphics[width=0.29\linewidth]{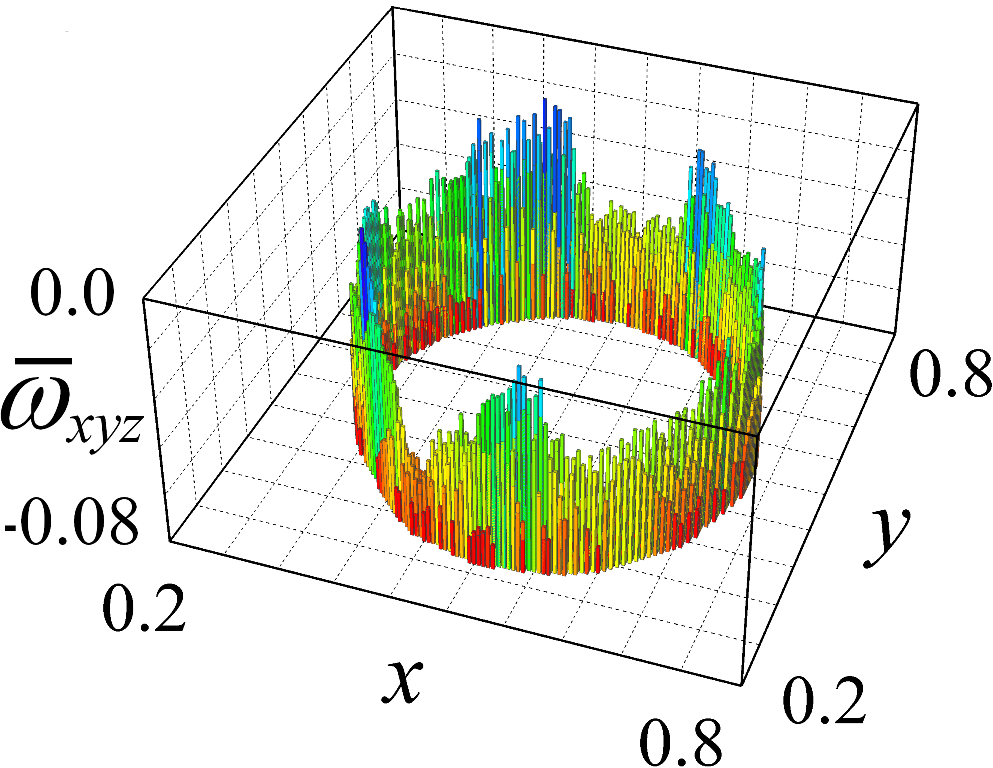}  \ 
  \includegraphics[width=0.04\linewidth]{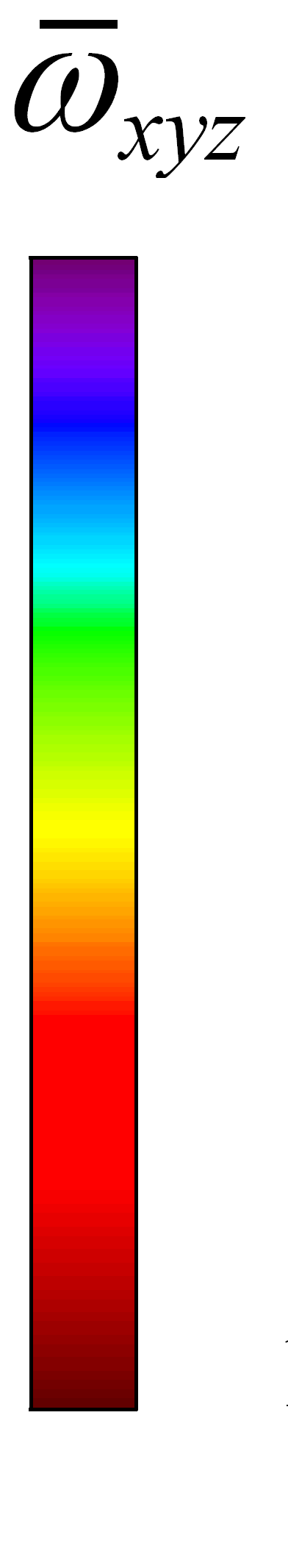}

  \includegraphics[width=0.29\linewidth]{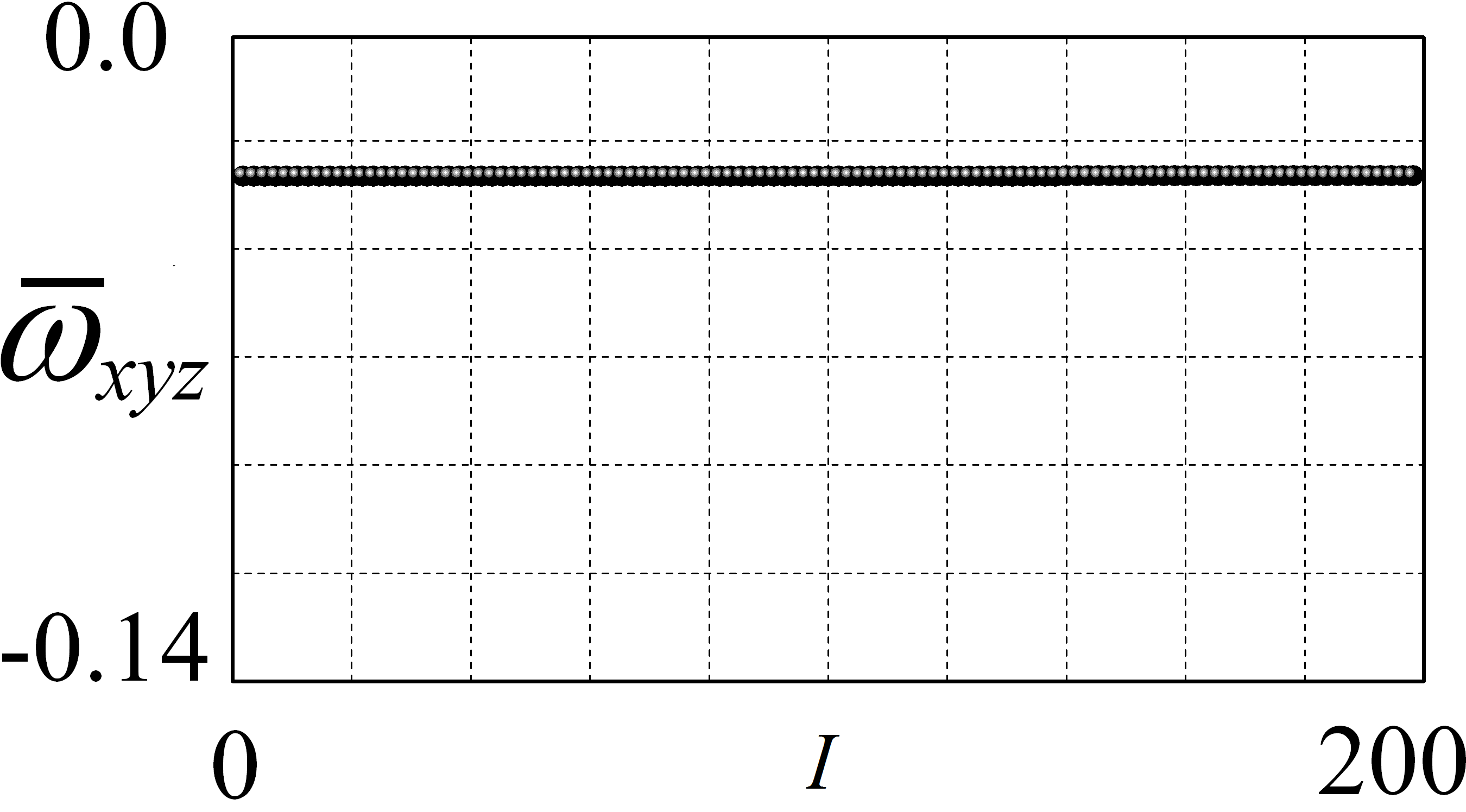}  \ 
  \includegraphics[width=0.29\linewidth]{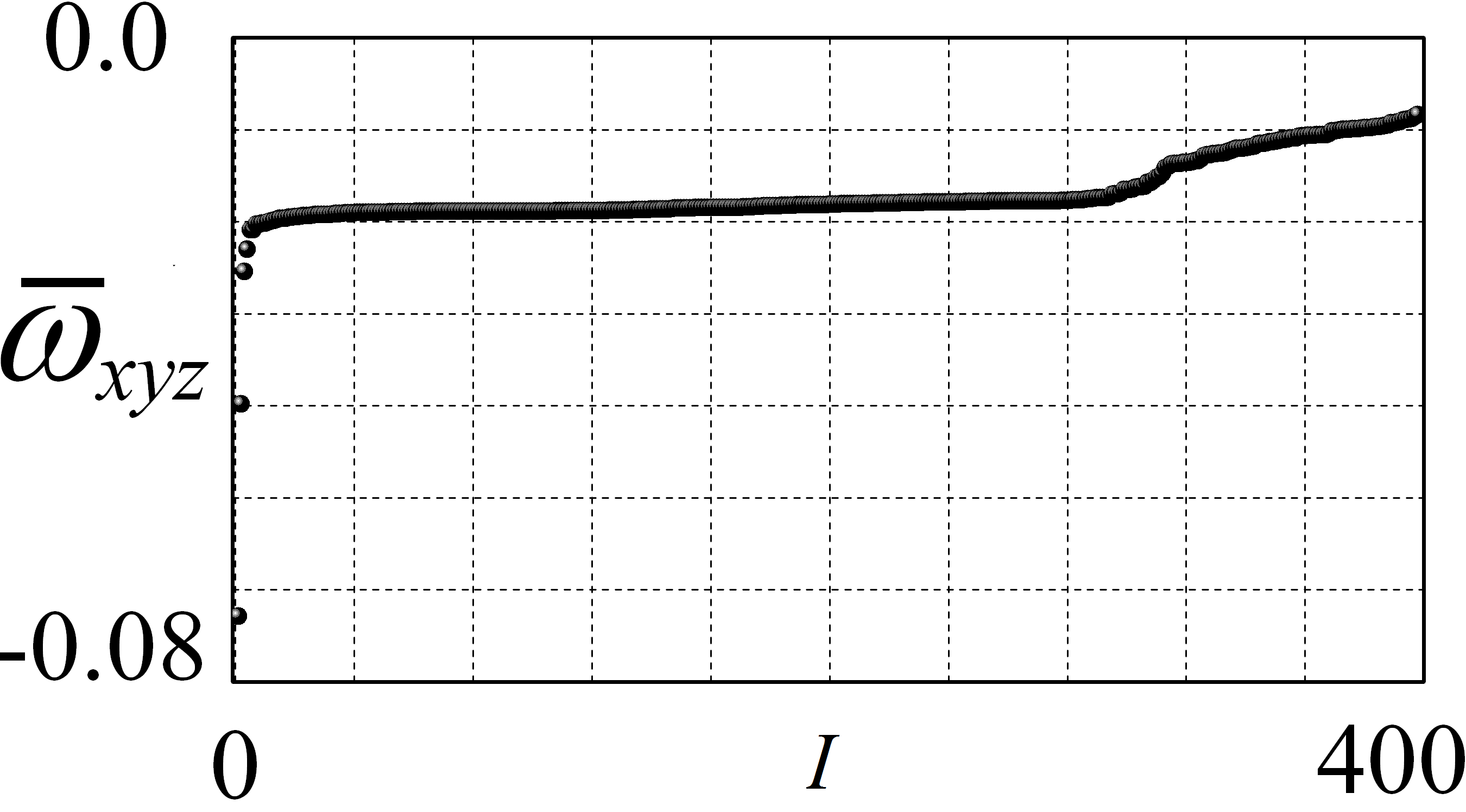}  \ 
  \includegraphics[width=0.29\linewidth]{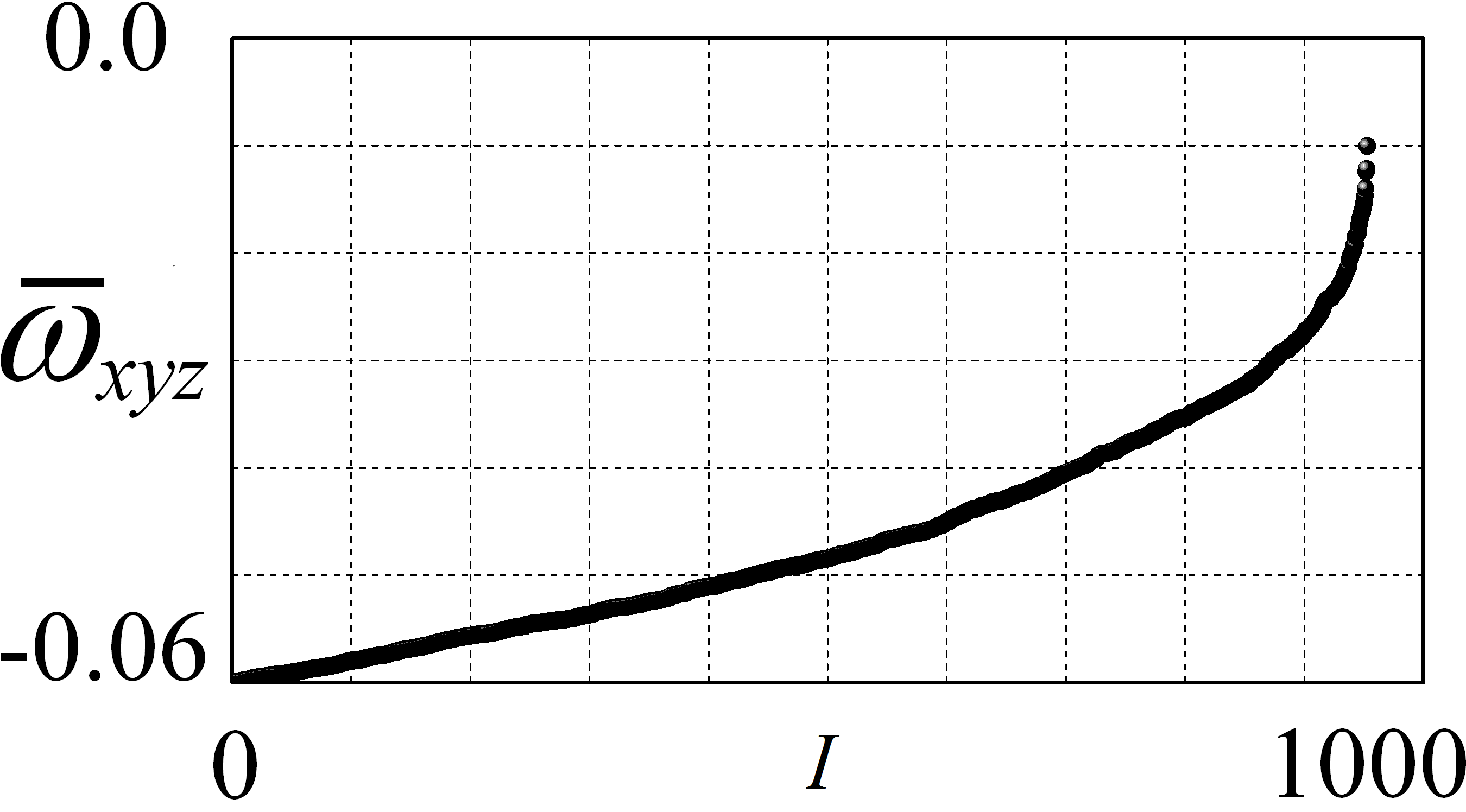}  \ 
  \includegraphics[width=0.04\linewidth]{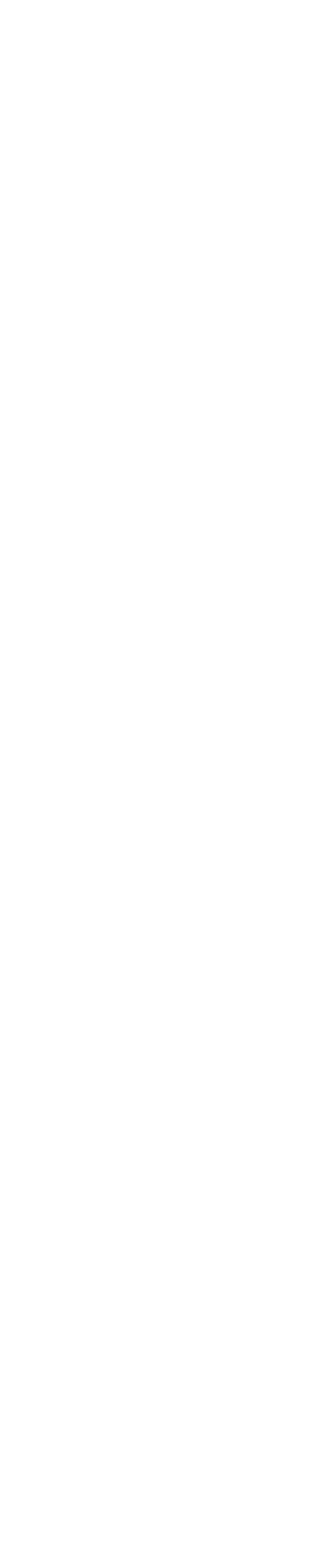}
 \hspace{3.5cm}} 

\hspace*{-2.8cm}
\begin{minipage}{1.15\textwidth}
 \caption{Examples  of  coherent, partially coherent, and  incoherent scroll ring chimeras.   Phase snapshots $\varphi_{xyz}$  (upper row), their phase cross-sections   along y=0.5,  the average frequencies  $\bar{\omega}_{xyz}$ (next two rows), ordered oscillator index (bottom row):
  (a) - coherent scroll ring chimera ($\alpha=0.35, \mu=0.02$), (b) - partially coherent scroll ring chimera ($\alpha=0.4, \mu=0.01$);  (c) - incoherent scroll ring chimera ($\alpha=0.5, \mu=0.007$). Common parameters  
   $r=0.01, \epsilon=0.05, N=200$. Simulation time $t=5\times10^4$, frequency averaging interval  $\Delta T = 10^4$.}
\end{minipage}
\label{fig:4}
\end{figure}

Scroll ring chimeras can have different inner structures  of filaments and different dynamics of their oscillators that are characterized by the time-averaged
frequencies  $\bar{\omega}_{xyz}$ of their oscillators.
The examples  of scroll ring chimeras with  coherent, partially coherent, and  incoherent inner parts with the major diameter of 0.5 are presented in Fig. 4.  Location of their parameter values in the parameter plane $(\alpha, \mu)$ for these examples are indicated by black points  a) - c) in Fig. 2.

In the case of a completely coherent scroll ring (Fig. 4(a)) all oscillators of the scroll ring chimera
rotate with the same time-averaged frequency   $\bar{\omega}_{xyz}$.
The scroll ring chimera with partially coherent oscillatory organization has a segments of the same average frequency of scroll rings oscillators (green color) (Fig. 4(b)). Its ordered oscillator index $I$ has a long band for oscillators with  the same average frequency $\bar{\omega}_{xyz}$.
In the case of a incoherent scroll ring chimera, its average frequencies profile   is chaotic   (Fig. 4(c)).
The dynamical complexity of the chimera state can be characterized by a number of positive Lyapunov
exponents. 

Lyapunov exponents  $\lambda_{n} $ characterize how the initial perturbations behave themselves along the whole trajectory and are defined as the eigenvalues of a matrix 
$$\Lambda = \lim\limits_{t \to \infty} \frac{1}{2t} \log(Y(t)Y^{T}(t)),$$ where the matrix  $Y$ is a solution of the linearized differential equation  $\dot{Y} (t) = J(t)Y(t)$  with the identity matrix as initial conditions, and  $J(t)$ is a Jacobi matrix of system (1). Lyapunov exponents were estimated using the QR approach \cite{djv2021}. 

We suggest that, in the considered example  (Fig. 4(c)), the scroll ring chimera has hyper-chaotic character with  the huge number, more than one hundred, of positive Lyapunov exponents.

\begin{figure}[ht!]
\begin{center}
\includegraphics[width=0.8\linewidth]{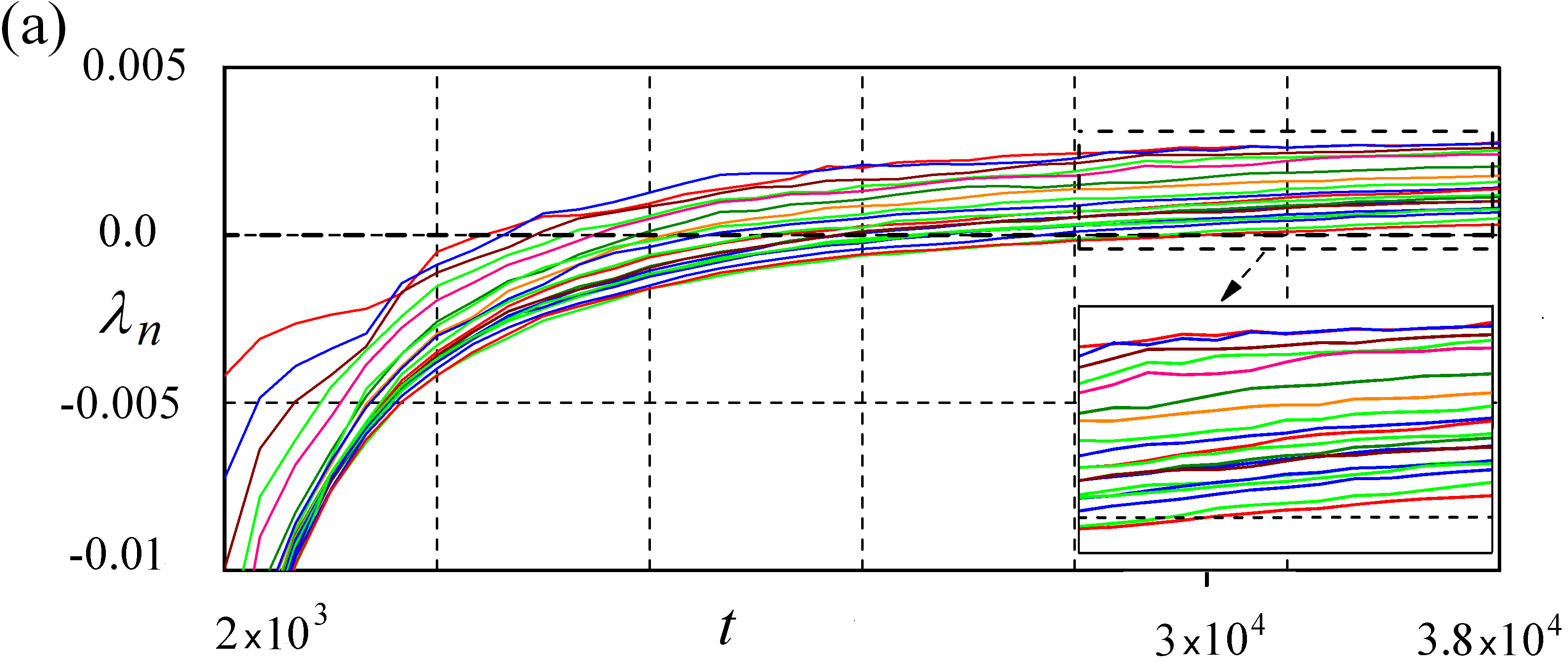}  \ \ \  \\ 
  \includegraphics[width=0.35\linewidth]{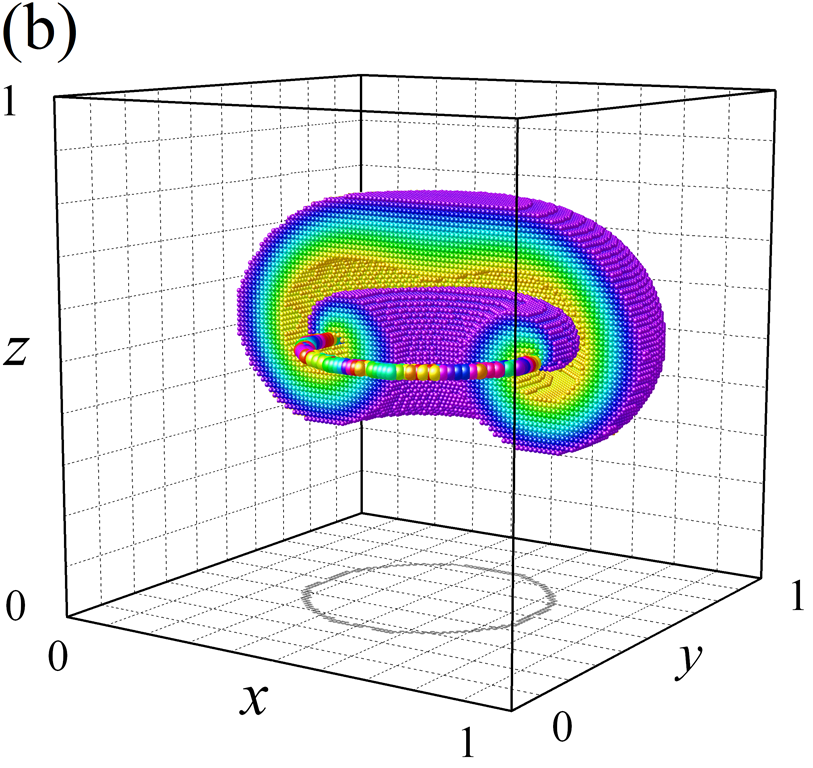}  \ \ 
 \includegraphics[width=0.045\linewidth]{Faza-insert-F4.png}
  \includegraphics[width=0.4\linewidth]{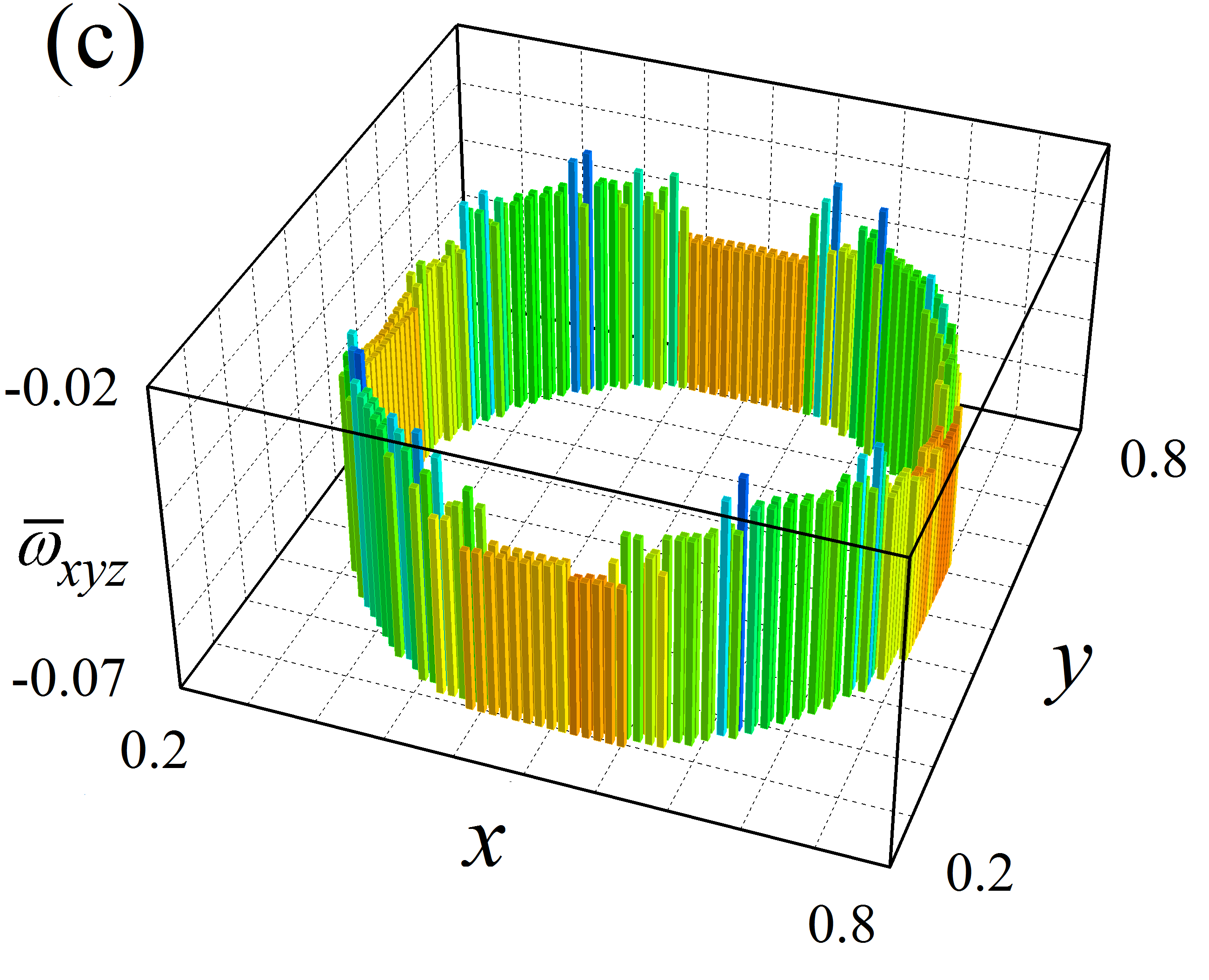}  \ \ 
  \includegraphics[width=0.04\linewidth]{AW-insert-Fig4.png}
\hspace*{-2.8cm}
\begin{minipage}{1.15\textwidth}
\caption { Lyapunov spectrum for 20 largest Lyapunov exponents (a) of the scroll ring chimera (b) and its average frequencies  $\bar{\omega}_{xyz}$ (c) for the parameters 
 $\alpha=0.45, \mu=0.01,  \epsilon=0.05, r=0.02, D=0.5, d=0.01, N=100$.  Simulation time $t=5\times 10^4$, frequency averaging interval  $\Delta T = 10^4$.  }
\end{minipage}
  \label{fig:5}
\end{center}
\end{figure}

\begin{figure*}[ht!]
 \center{
  \includegraphics[width=0.2\linewidth]{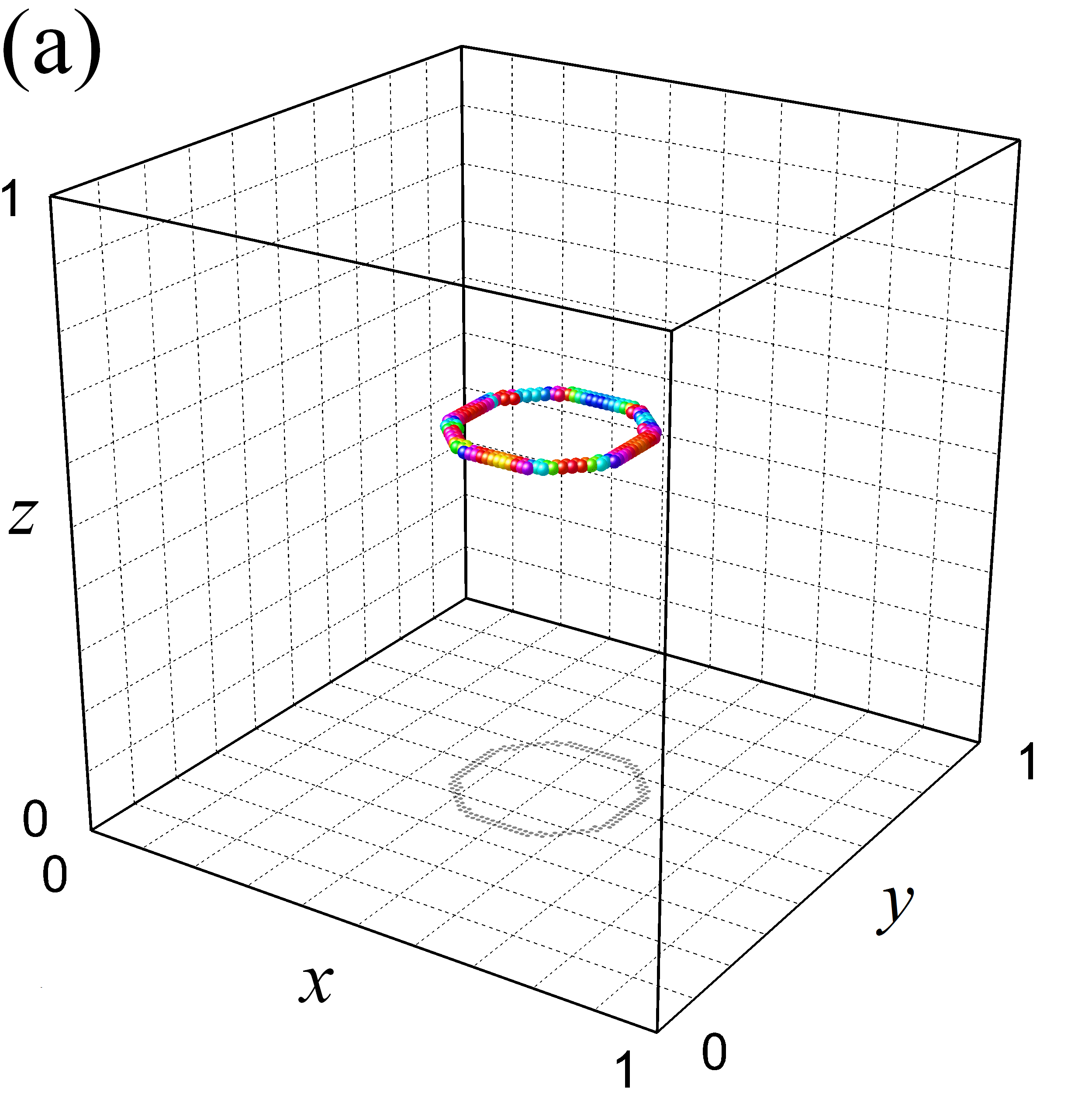}  \ \   \ 
  \includegraphics[width=0.2\linewidth]{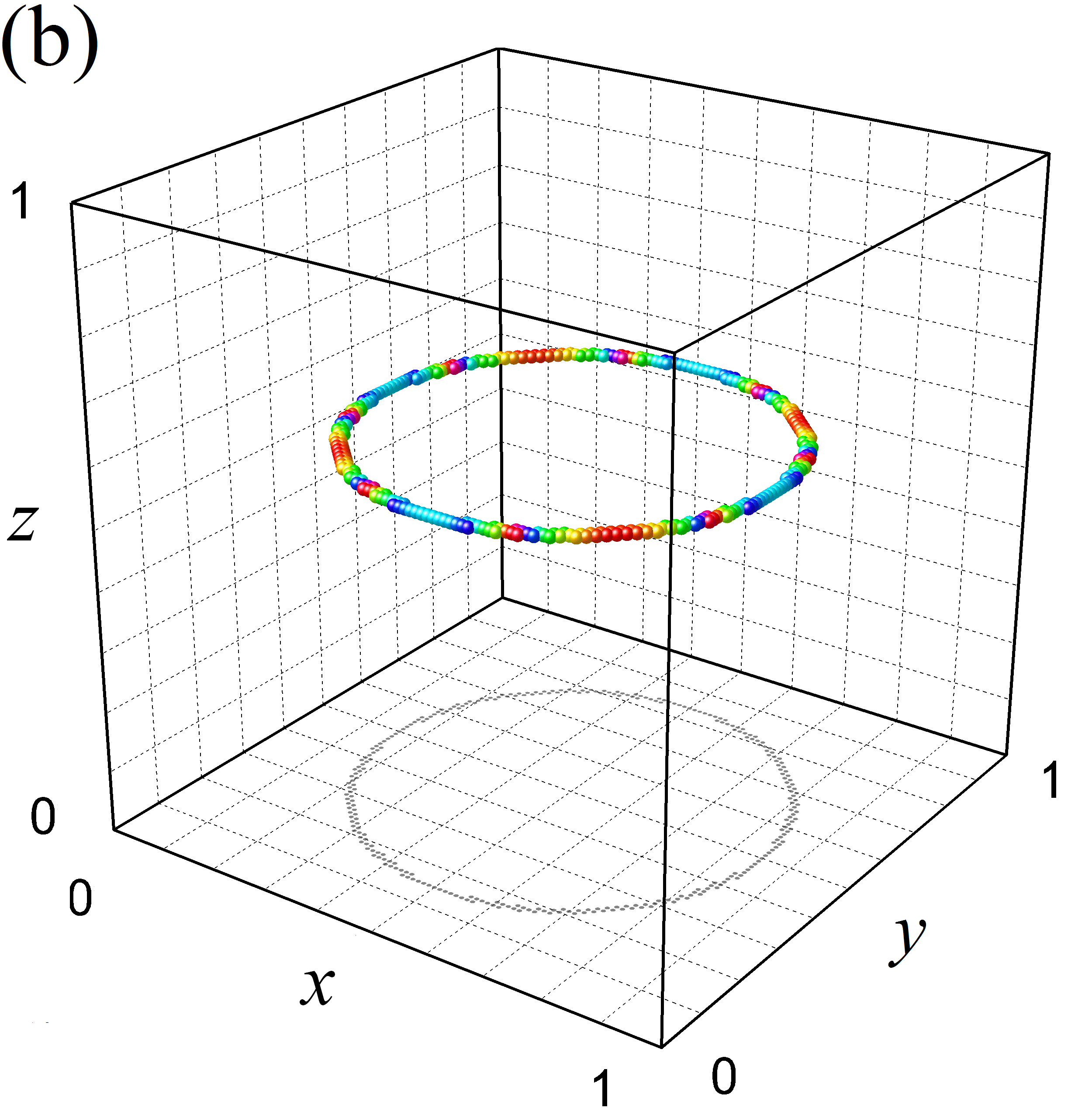} \   \  \
  \includegraphics[width=0.2\linewidth]{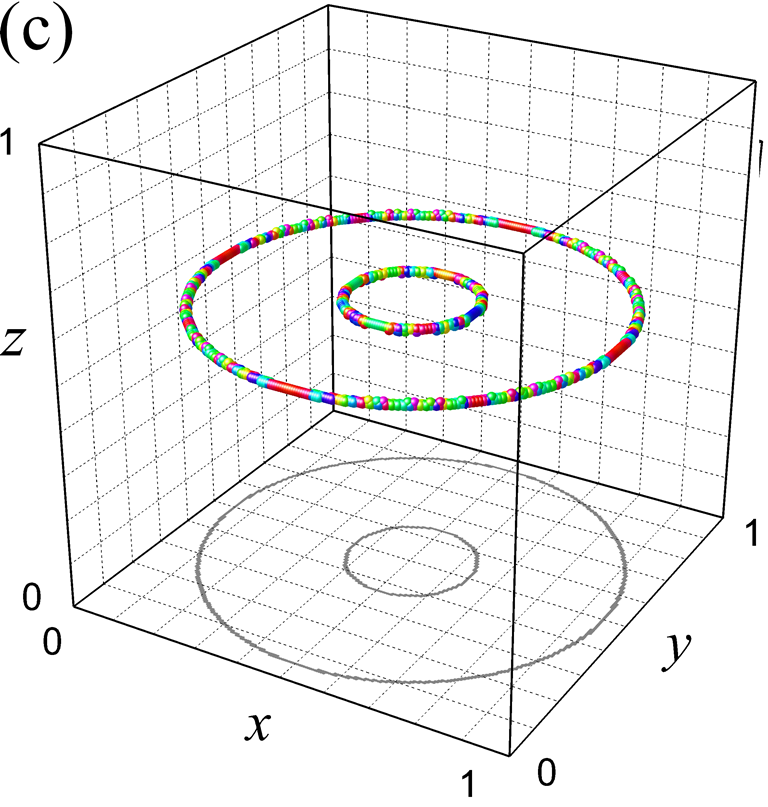} \ \   \
  \includegraphics[width=0.2\linewidth]{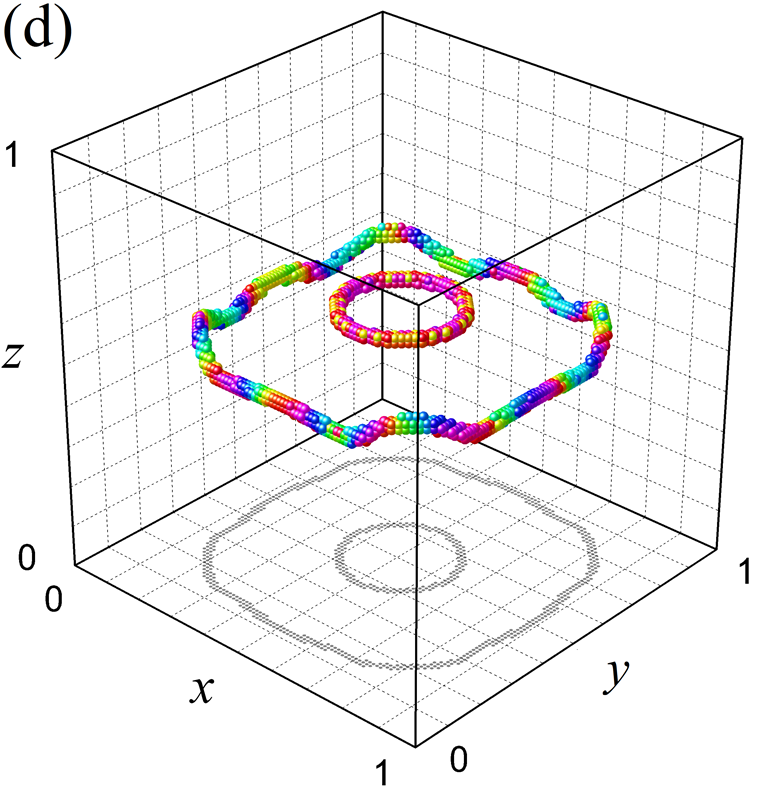} \ \
  \includegraphics[width=0.03\linewidth]{Faza-insert-F4.png}

\includegraphics[width=0.19\linewidth]{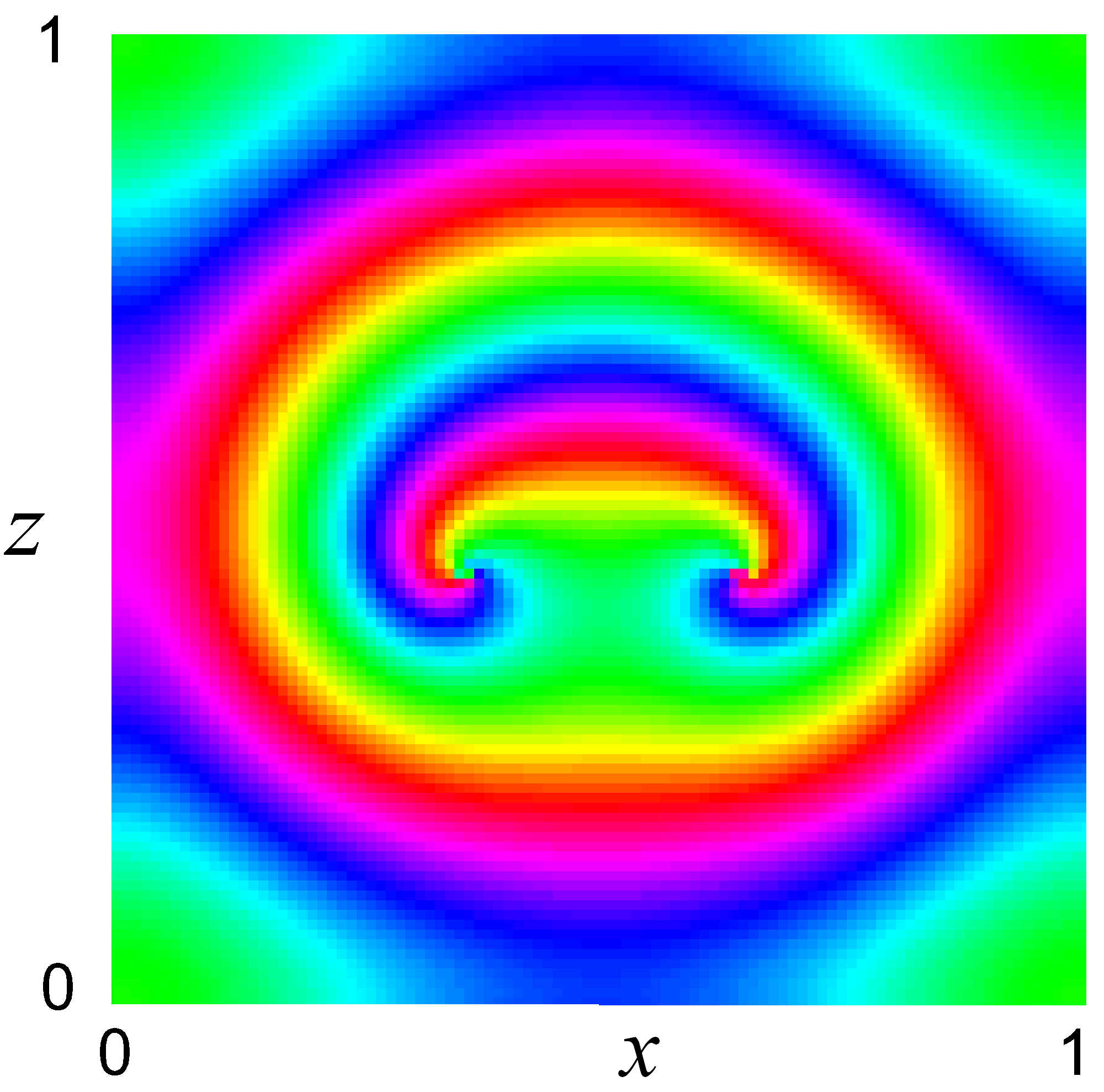}   \   \  \
\includegraphics[width=0.2\linewidth]{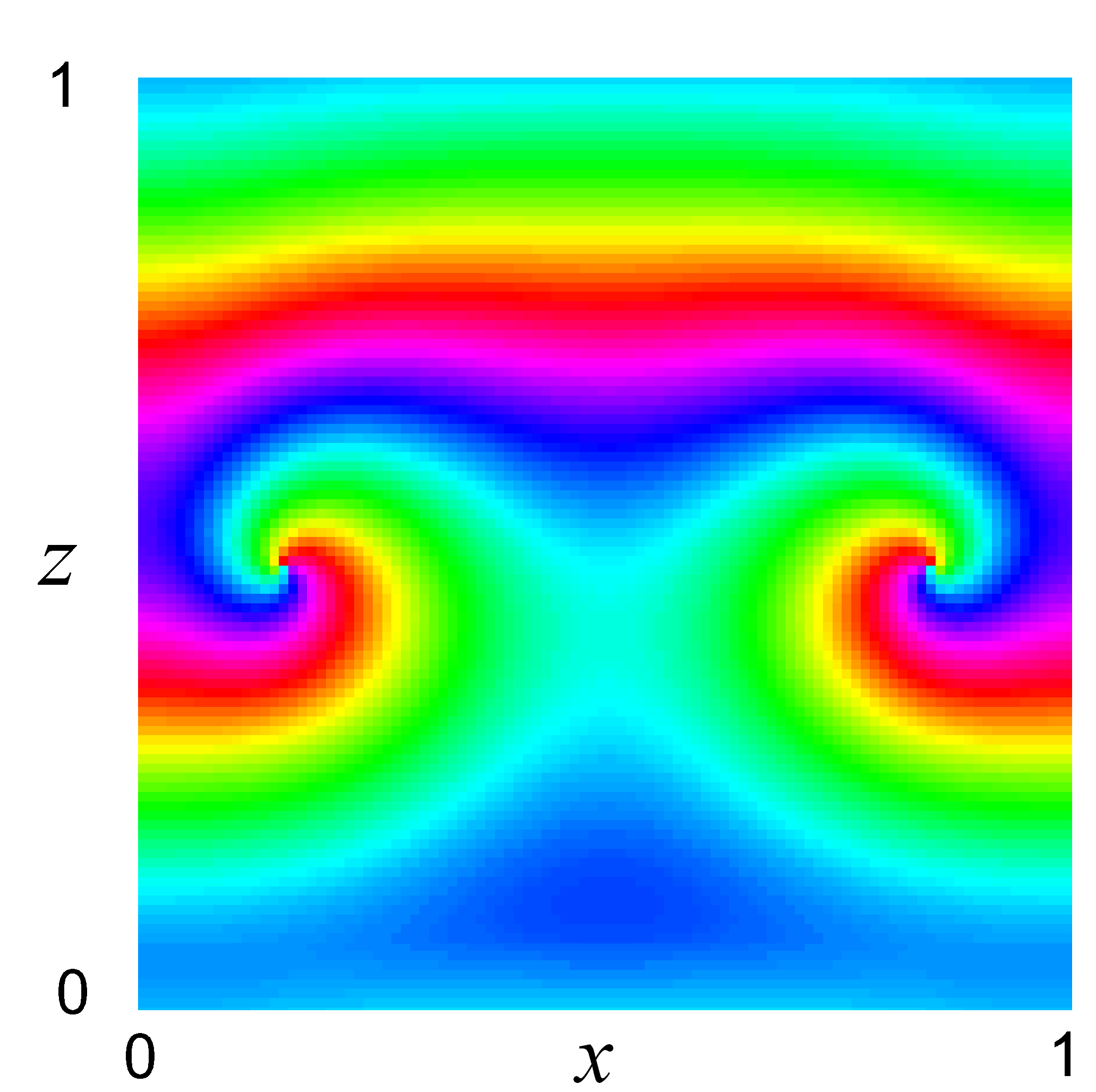}  \   \   \
\includegraphics[width=0.2\linewidth]{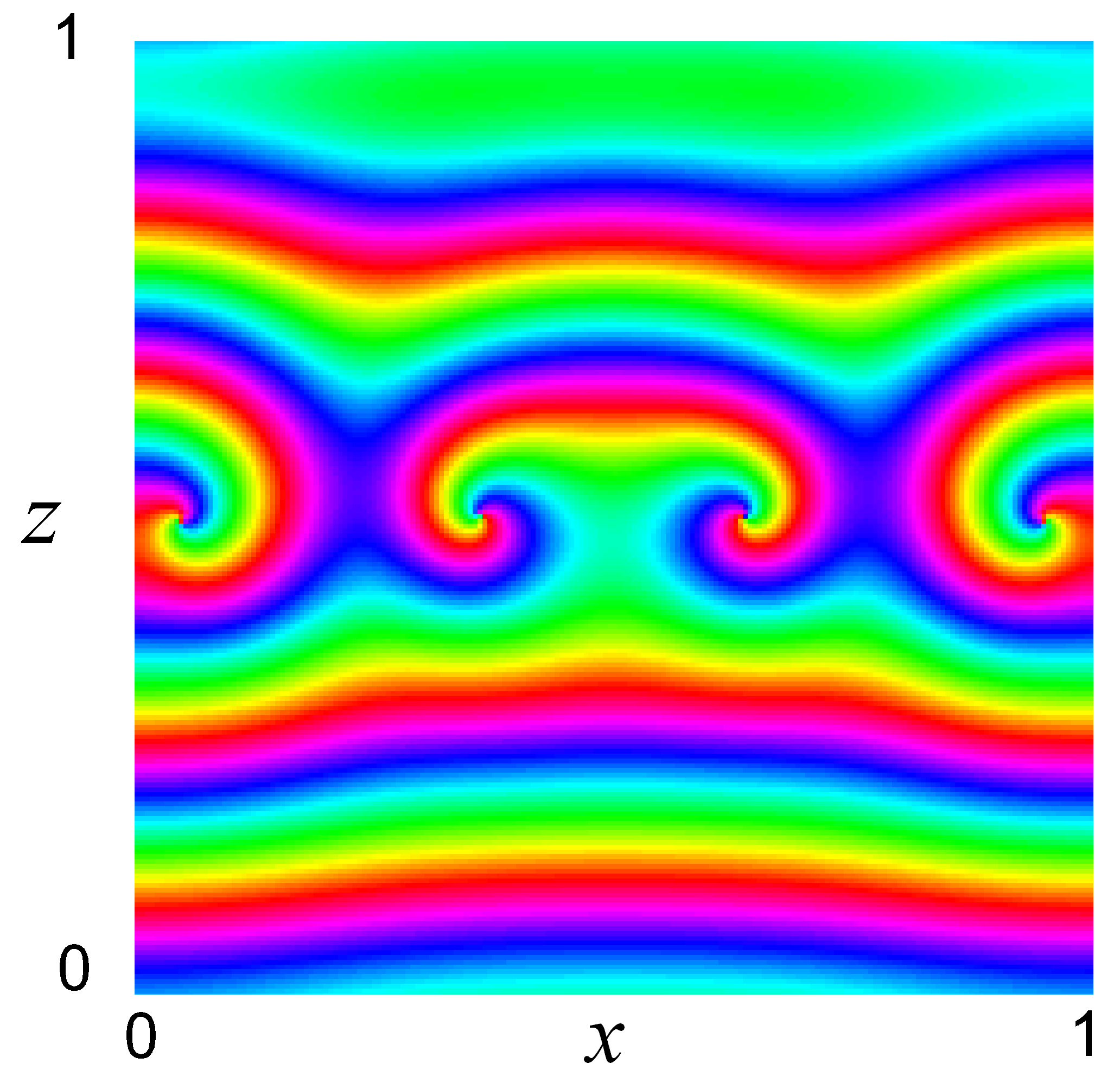}  \  \  \  \
\includegraphics[width=0.21\linewidth]{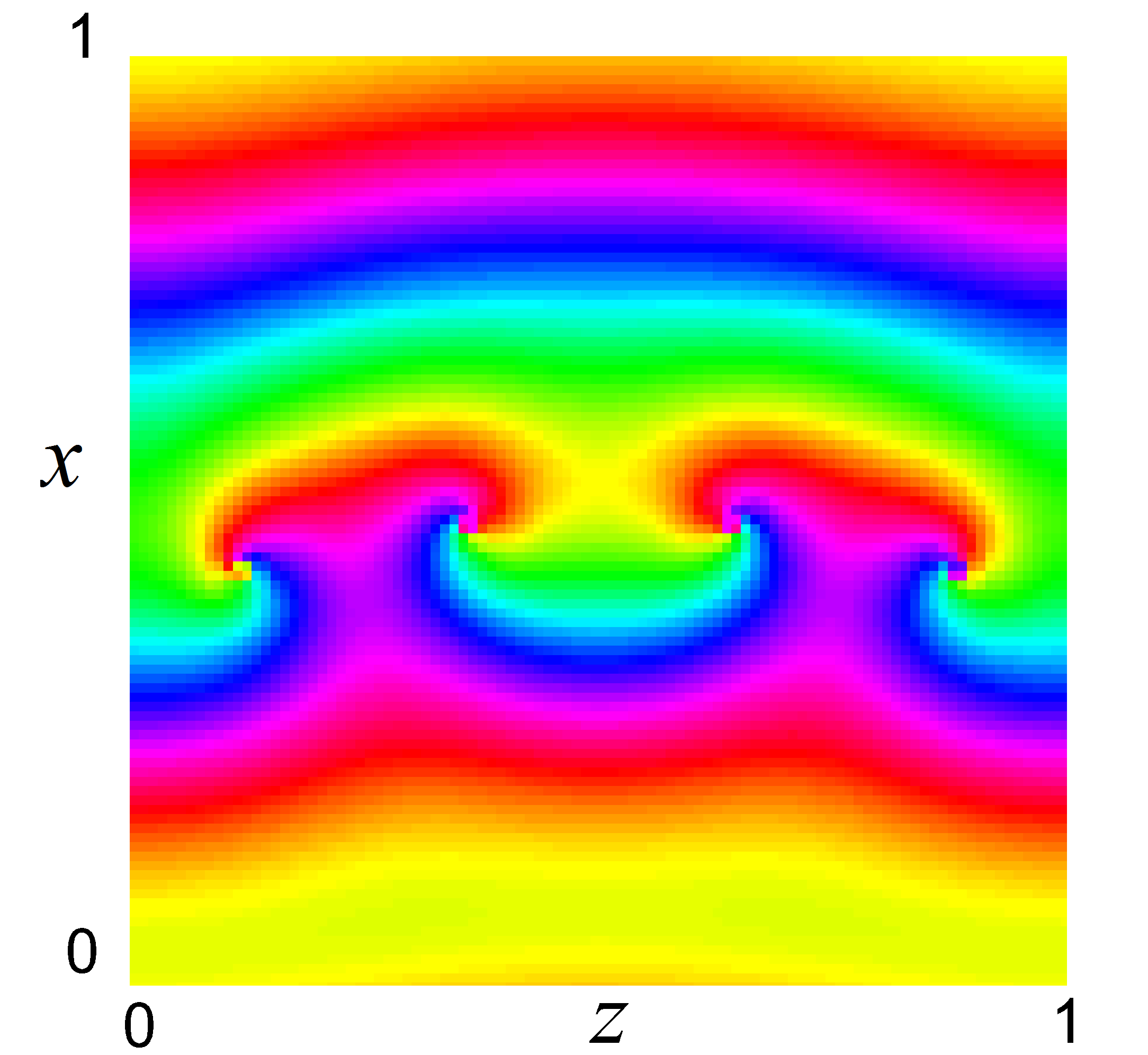}   \  \
  \includegraphics[width=0.03\linewidth]{Faza-insert-F4.png}  \  

\vspace*{0.2cm}

  \includegraphics[width=0.2\linewidth]{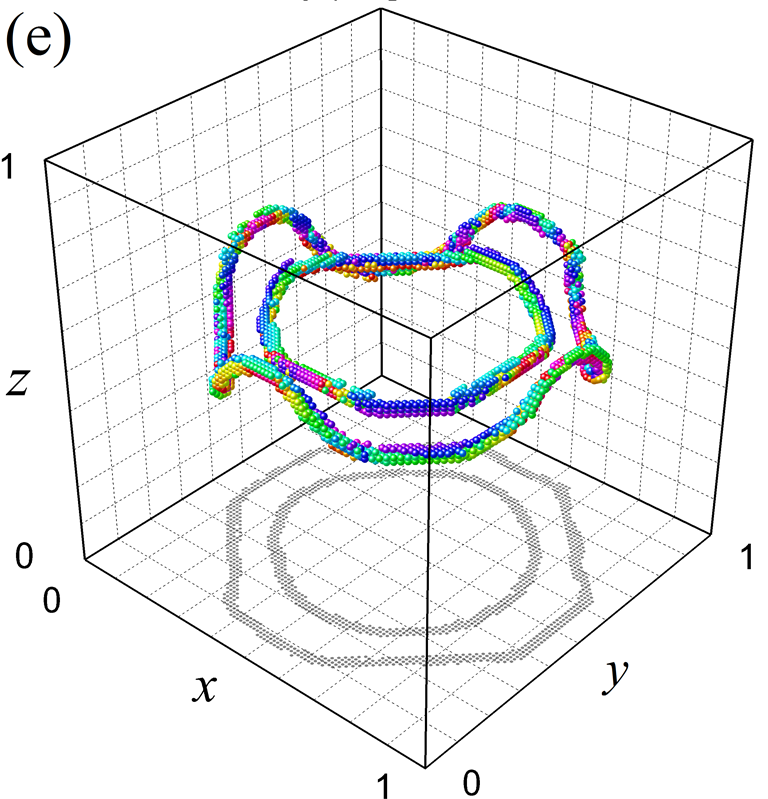} \ \  \
  \includegraphics[width=0.2\linewidth]{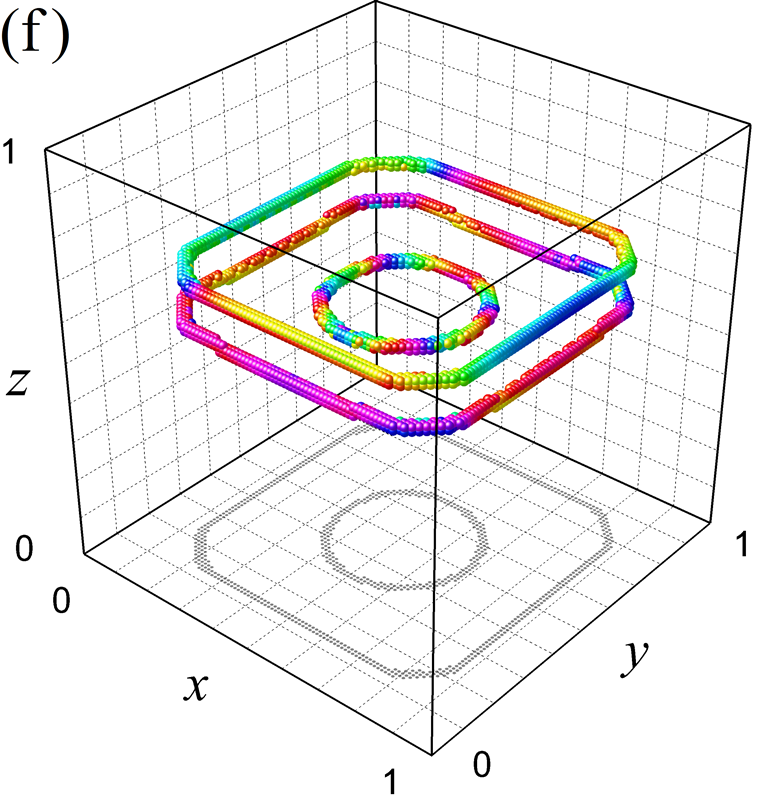}  \ \ \
  \includegraphics[width=0.2\linewidth]{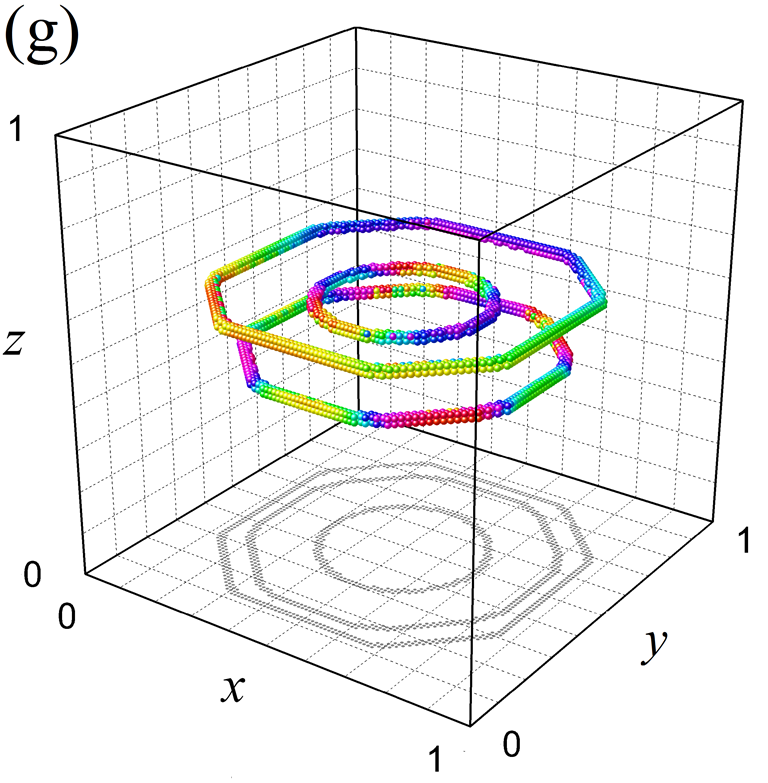}  \ \ \
  \includegraphics[width=0.2\linewidth]{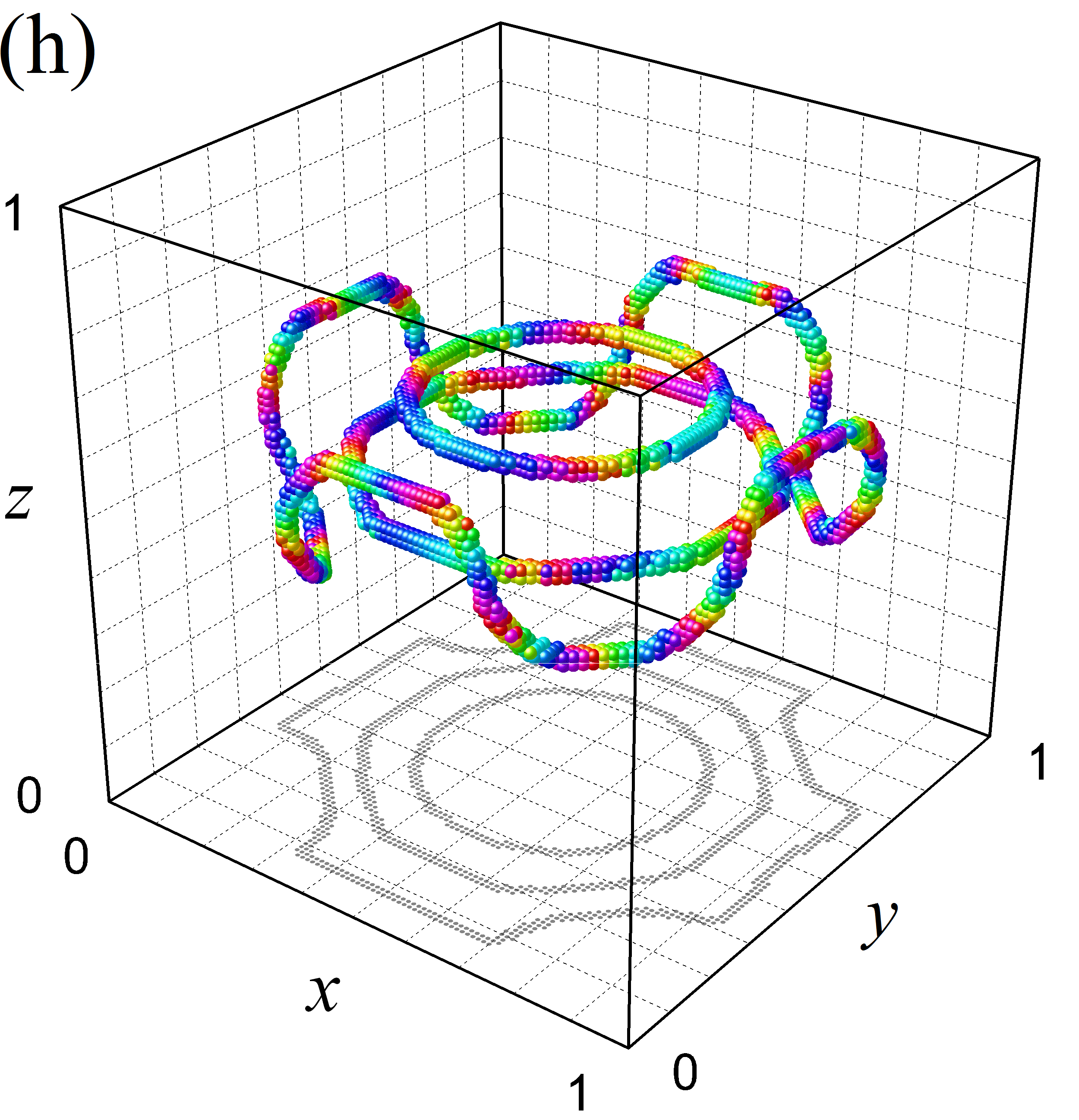}  \  
  \includegraphics[width=0.03\linewidth]{Faza-insert-F4.png}

\includegraphics[width=0.2\linewidth]{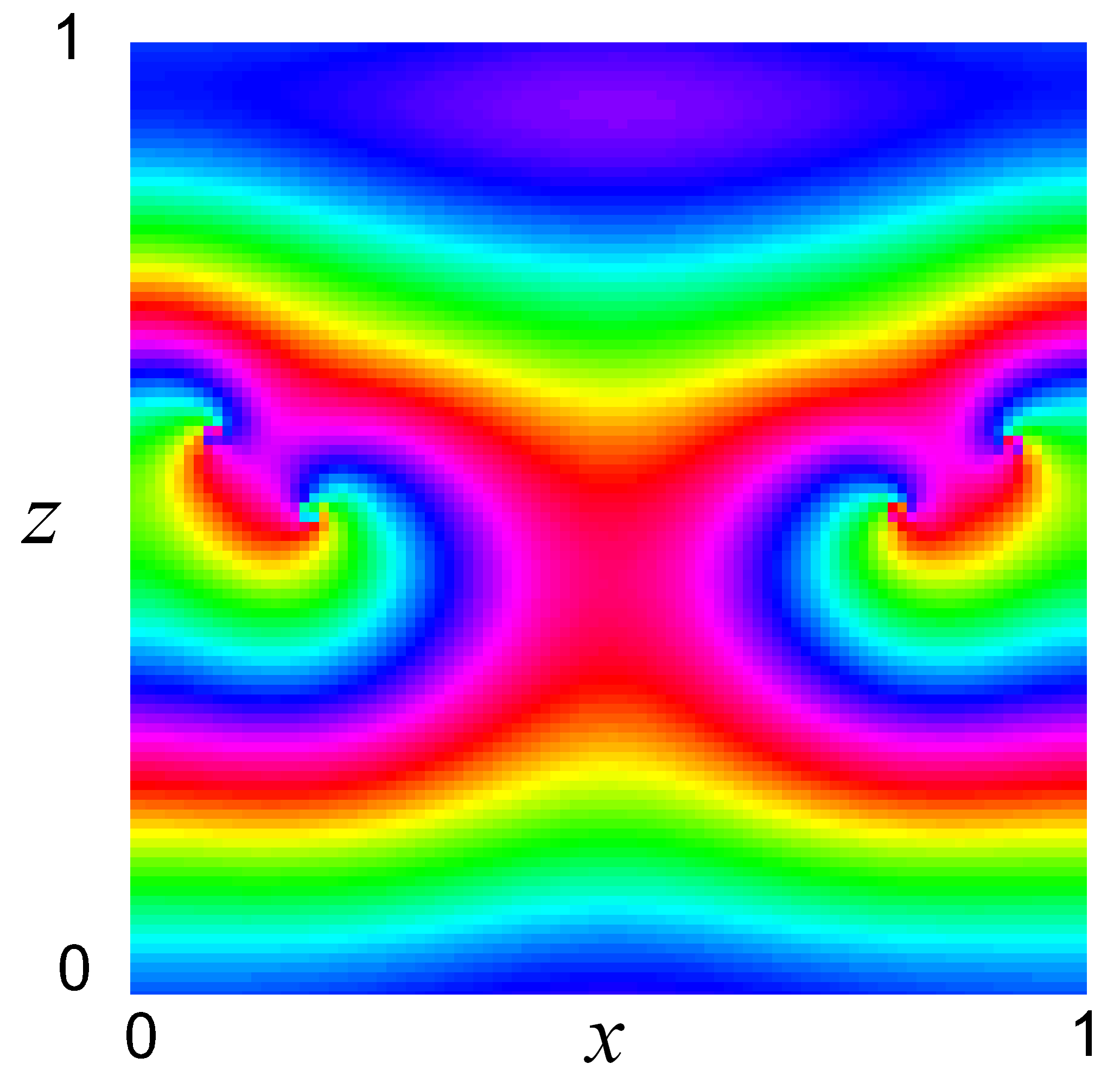}  \ \  \
\includegraphics[width=0.2\linewidth]{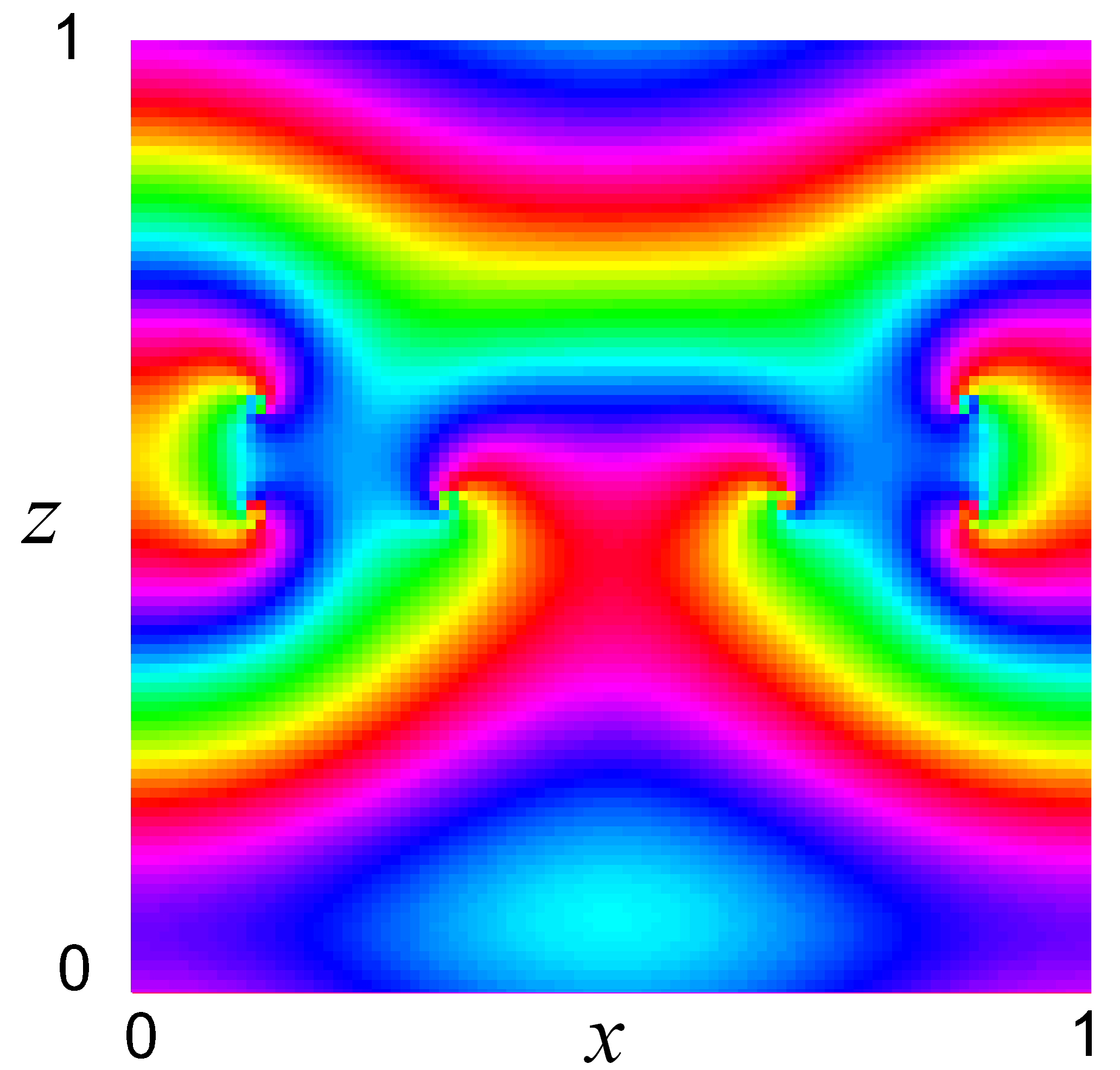} \ \  \
\includegraphics[width=0.2\linewidth]{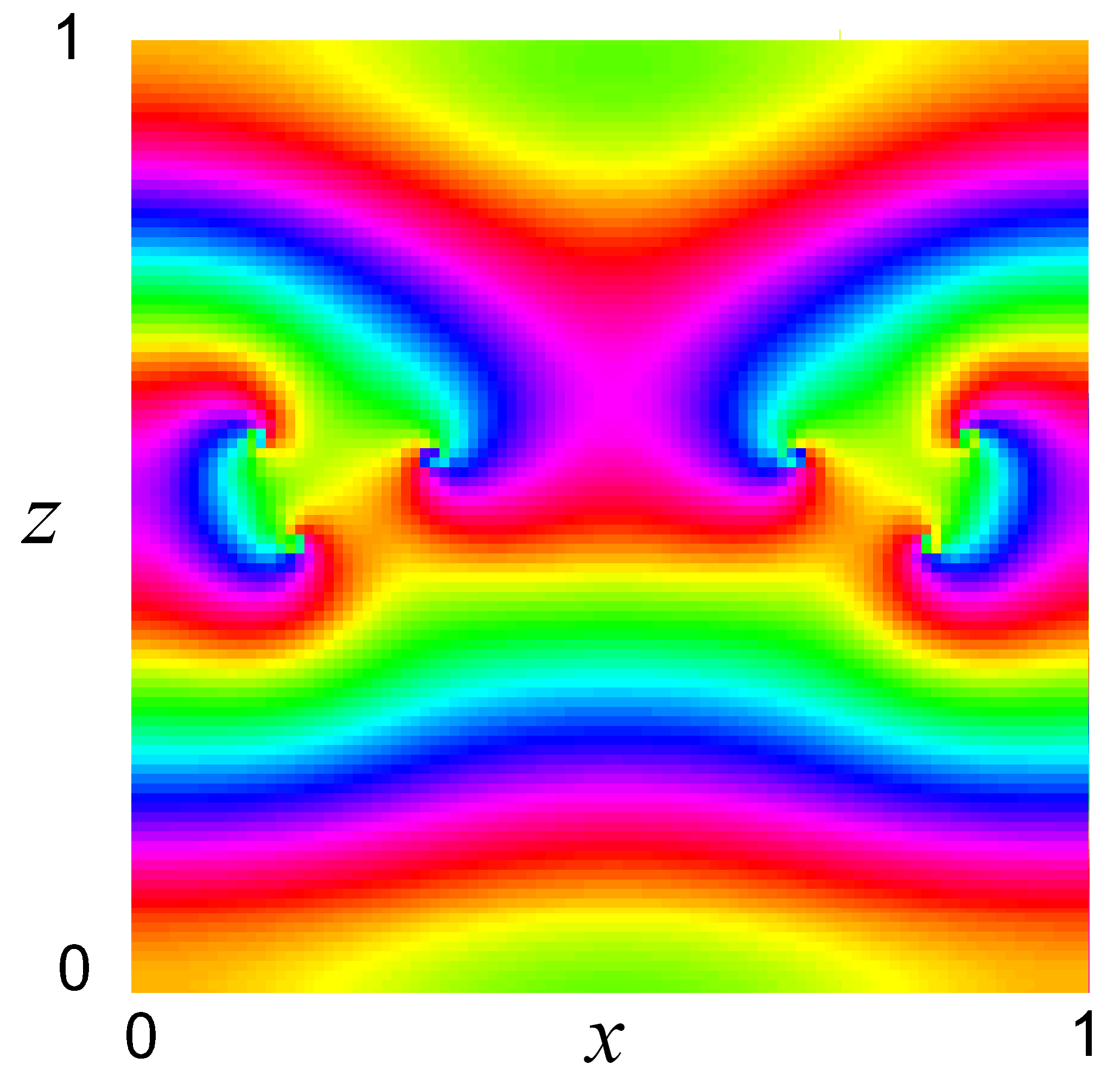} \ \  \
   \includegraphics[width=0.2\linewidth]{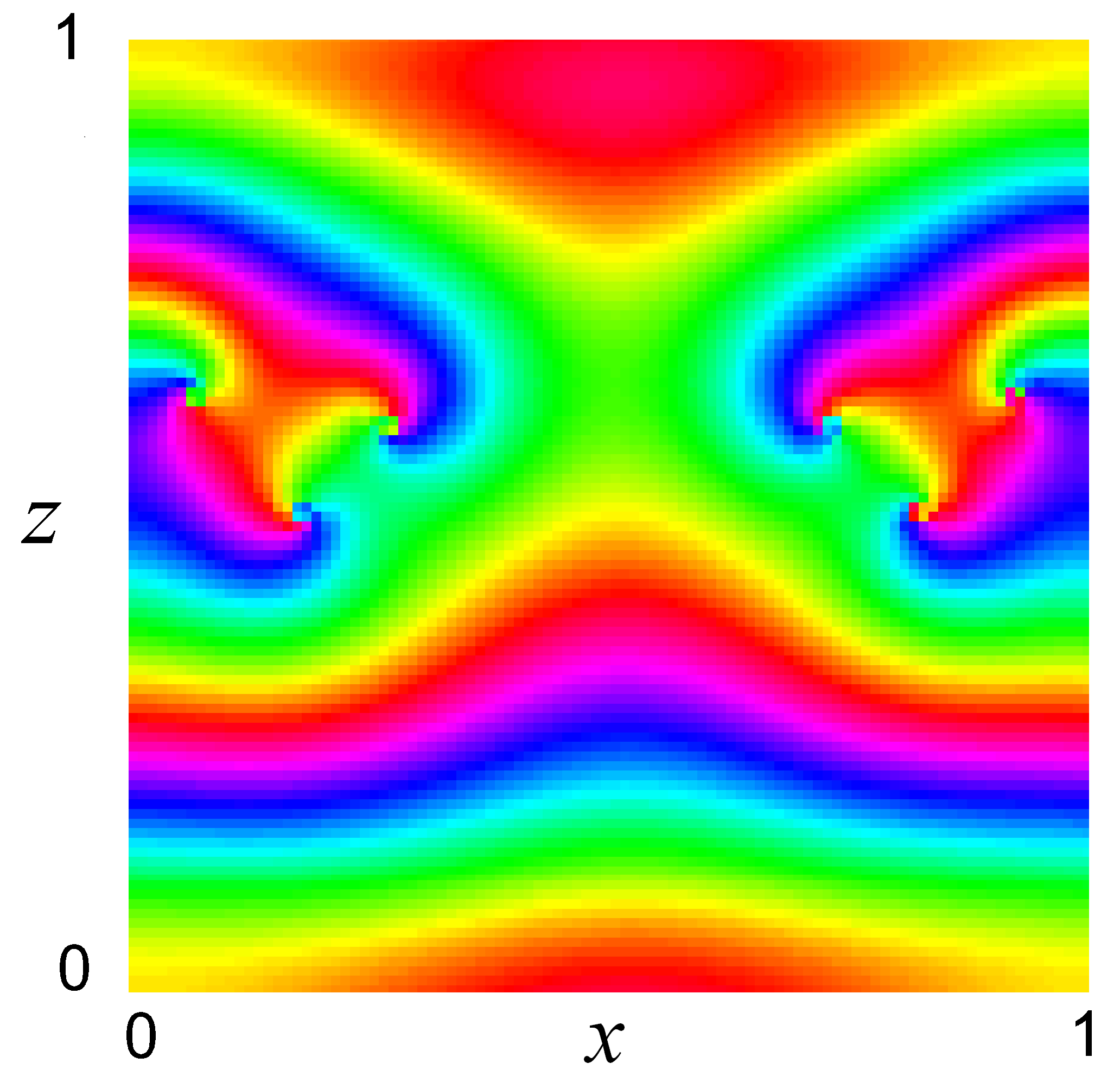} \   \ 
   \includegraphics[width=0.03\linewidth]{Faza-insert-F4.png}  }
\hspace*{-2.8cm}
\begin{minipage}{1.15\textwidth}
 \caption{ Examples of scroll ring chimeras generated by the  initial conditions (2), (3). 
Phase distributions  $\varphi_{xyz}$ and their cross-sections along $y = 0.5$:
  (a) - scroll ring with a major diameter of $0.3$ ($r=0.02, D=0.3, d=0.01, N=100$), (b) -  scroll ring with a major diameter of $0.7$ ($r=0.02, D=0.7, d=0.01, N=100$),  (c) -  double concentric scroll rings ($r=0.02, D=0.9, d=0.005, N=200$),  (d) -  double scroll rings with wave profile ($r=0.03, D=0.88, d=0.1, N=100$),  (e) -  double scroll rings with wave profile ($r=0.03, D=0.58, d=0.2, N=100$),  (f) -  three scroll rings ($r=0.03, D=0.66, d=0.2, N=100$), (g) -  three scroll rings ($ r=0.03, D=0.64, d=0.22, N=100$), 
(h) -  three scroll rings with wave profile $(r=0.03, D=0.74, d=0.24, N=100$). Common parameters   $\alpha=0.38, \mu=0.02,  \epsilon=0.05$.
 Simulation time $t=5\times10^4$. Coordinates  $x=i/N,  y=j/N,  z=k/N$. } 
\end{minipage}
\label{fig:6}
\end{figure*}

 Figure 5 illustrates  the Lyapunov spectrum of  20 largest Lyapunov exponents (a) of the scroll ring chimera (b) and its average frequencies  $\bar{\omega}_{xyz}$ (c). At the time $t \approx 3\times10^4$, all 20 largest Lyapunov exponents become positive, confirming the hyper-chaotic character of this scroll ring chimera state.

We believe that if system (1) with dimension $N$ increases, then it will be possible, using the   initial conditions (2) and (3),  to obtain different scroll ring  chimeras including such ones with a stepwise profile with the average frequencies  $\bar{\omega}_{xyz}$ similarly to the structure of spiral cores in the  2D Kuramoto model with inertia   \cite{msm2020}.

The  initial conditions (2) and (3) can generate  single scroll ring or scroll toroid chimeras, as well  as multiple scroll pattens with different major and minor diameters. Some filaments of them are transformed into polygons or can have a wave profile in the $z$ direction.

In Fig. 6, a few examples of scroll ring chimeras, as a  result  of the simulation with the use of the  initial conditions (2) and (3), are presented 
for the fixed parameters  $\alpha=0.38, \mu=0.02,  \epsilon=0.05$. The initial conditions are taken
 in the form of (2) for Fig. 6(a, b, e, h) and in the form of (3) for Fig. 6(c, d, f, g).

We note that the proposed  initial conditions (2), (3) can be used  for any values of $N$ of system (1).  We present the results of their application in the cases of $N = 100$ and $N = 200$ in Fig. 6(a, b, d-h) and Fig. 6(c), respectively.

In addition to the example of a scroll ring chimera with the major diameter of 0.5 (Fig 3(a)), the examples of  scroll ring chimeras  with the major diameter of 0.3 and 0.7 are presented in Fig. 6(a) ($ D=0.3, d=0.01$)  and Fig. 6(b) ($ D=0.7, d=0.01$), respectively.

The example of a multiple scroll ring chimera obtained from the  proposed initial conditions is given in 
Fig. 6(c-h):
(c) -  double concentric scroll rings ($D=0.9, d=0.005$),  (d) -  double scroll rings with wave profile ($D=0.88, d=0.1$), (e) -  double scroll rings with wave profile ($D=0.58, d=0.2$), (f) -  three scroll rings ($D=0.66, d=0.2$), (g) -  three scroll rings with polygon profile ($D=0.64, d=0.22$),  (h) -  three scroll rings  with wave profile ($D=0.74, d=0.24$).  Multiple scroll pattens in the form of polygons or with wave profile  can be generated by initial conditions (2), (3) with  parameter $d > 0.09$.
So, using the  initial conditions (2), (3)  with different parameters $D$ and $d$, we can obtain as many multiple scroll wave ring and toroid chimeras as we simulate. 

Our simulation shows that all scroll ring chimeras obtained in this way are  stable with respect to perturbations of the phase variables  $\varphi_{xyz}$ and frequencies $\omega_{xyz}$ by a uniformly distributed noise.  The perturbed scroll ring chimeras still exist and retain their shapes under a perturbation of the 
amplitude less than 0.5 at the parameters $\alpha=0.38, r=0.02, \mu=0.1,  \epsilon=0.05$, and $N=200$. Stronger perturbations lead to changing the shapes of scroll ring chimeras, their destruction with complete oscillatory synchronization, or the creation of different types of scroll wave chimera states. Moreover, if the disturbed scroll ring lies in the parameter region, where solitary states exist,  the perturbation can give rise to another scroll ring chimera with solitary clouds \cite{mso2020}. Due to these  properties, we can build complex patterns in the system (1), for example, consisting of scroll ring chimeras
and other patterns.

\begin{figure}[ht!]
\centering  
\includegraphics[width=0.37\linewidth]{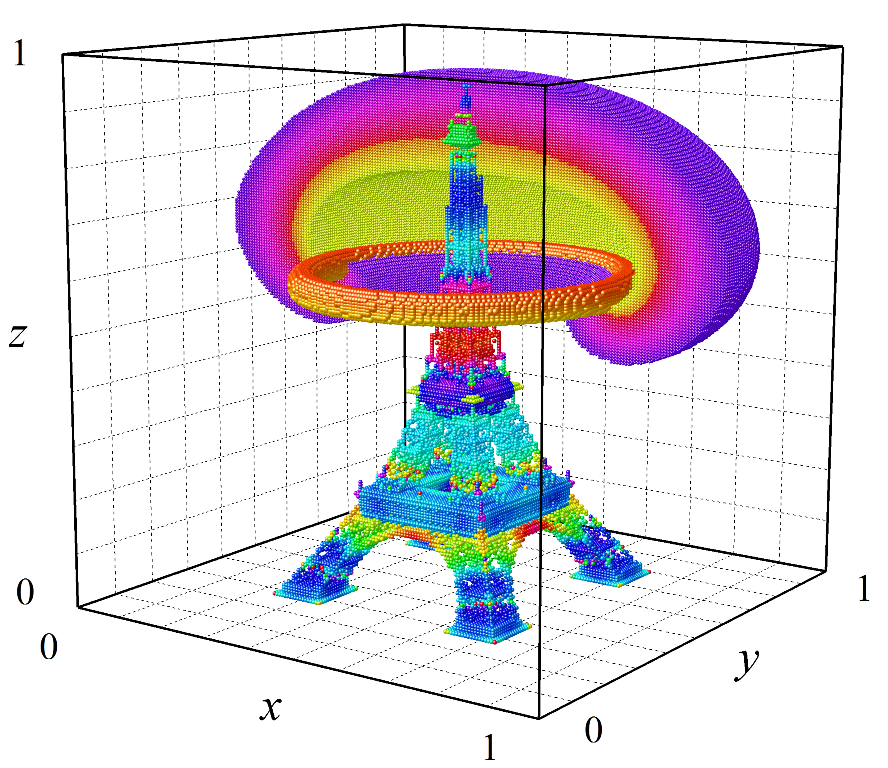} \     \ \ 
  \includegraphics[width=0.045\linewidth]{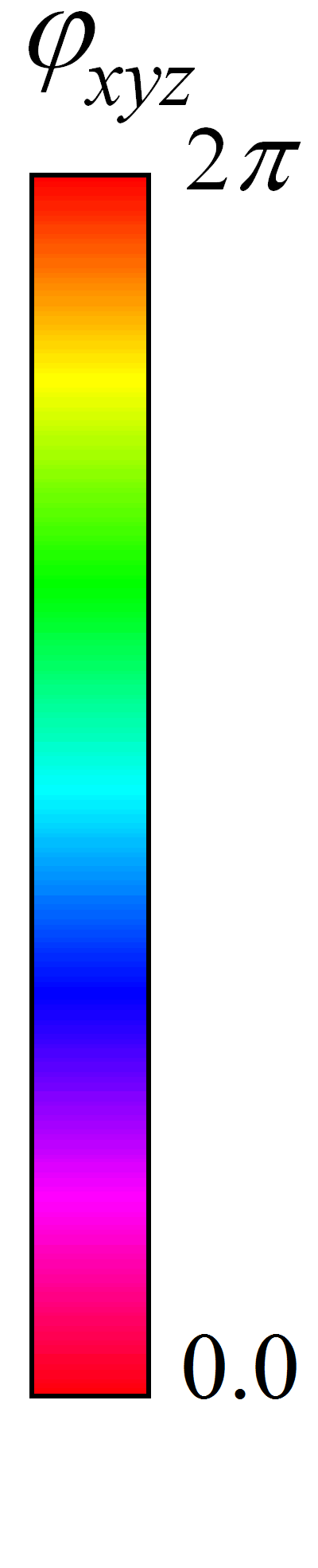}
\hspace*{-2.4cm}
\begin{minipage}{1.15\textwidth}
 \caption{Phase snapshot of the 3D image of the Eiffel Tower  with a scroll ring chimera. The parameters $ \alpha=0.4, r=0.04, \mu=0.1,  \epsilon=0.05, D=0.5, d= 0.005, N=200$. A half of the wave has been removed to permit
the view of the scroll ring chimera and  the Eiffel Tower image. The space–time dynamics of the chimera states are illustrated by video in supplemental data.
}
\end{minipage}
\end{figure}

Finally, to illustrate the possibility of the coexistence  of the scroll rings with other patterns in the 3D oscillatory network, we present the Eiffel tower image surrounded by scroll ring chimeras in  Fig. 7.
The initial conditions for the patterns  were built, by using the Eiffel tower solitary oscillators  and the  initial conditions (2) for a scroll ring chimera with the parameters $D=0.5$ and $d= 0.005$ placed in a 3D cube at $z=0.6$. 
Distribution of scroll ring chimera oscillators was taken as the base for initial conditions. 
Then the model of Eiffel tower for the 3D printing was used as a spatial
template for the placement of solitary oscillators inside a scroll ring chimera. Model for 3D printing contains spatial coordinates  of voxels  that should be filled with plastic. These voxel coordinates were scaled to the size of the scroll ring chimera system and  used for placement of solitary oscillators inside a system with a scroll ring chimera. 
The average frequency of solitary oscillators was selected to be close to the average frequency of scroll ring oscillators. Due to properties of solitary states  \cite{mso2020}, there is no interference between solitary oscillators of the Eiffel tower and the oscillators of a scroll ring chimera after several hundreds time units transient.
This image is a stable solution of the 3D Kuramoto model with inertia (1). 

\vspace*{0.4cm}

The scroll ring has a ring-like filament and is related to a scroll wave and a toroidal vortex/swirl as a special case. Scroll rings also appear in a variety of experimentally observed effects and as solutions of the equations of realistic mathematical models. Equations similar to (1) appear in different problems of hydrodynamics, aerodynamics, and plasma dynamics and are related to flows, flames,   superfluidity etc. (see, e.g.,   \cite{dsp2014,hy2005, gmf2020}). The fibrillation of  
cardiac tissue is described by scroll ring dynamics,  and some chemical reactions lead to the propagation of scroll ring reagents' concentration waves.  
The appearance of  scroll rings and other related phenomena are sometimes considered as harmful (flow turbulence, fibrillation)  and, sometimes, as useful (combustion chambers, chemical reactions). Knowledge of  conditions, where scroll rings can and cannot exist, may be inculcated in practice: design of new devices, therapies, technologies.

\section*{Acknowledgments} 

The authors are grateful to 
the Ukrainian Grid Infrastructure for providing the computing
cluster resources and the parallel and distributed software used
during this work.

\end{document}